\documentclass[11pt]{article}
\pdfoutput=1
\usepackage{jheppub}

\usepackage{color}
\usepackage{amsmath}
\usepackage{pifont}
\usepackage{bbold}

\usepackage{bbm}
\usepackage{verbatim}   
\usepackage{subfigure}
\usepackage{acronym}

\usepackage{amsfonts}
\usepackage{amssymb}
\usepackage{mathrsfs}
\usepackage{graphicx}
\usepackage{multirow}
 \usepackage{slashed}

\graphicspath{{./fig/}}

\newcommand{\ie}{{\it i.e.}}          
\newcommand{\eg}{{\it e.g.}}          \newcommand{\etc}{{\it etc.}}
                   
\newcommand{\kahler}{K\"{a}hler }          

\newcommand{\lang}{\mathcal{L}}
\newcommand{\N}{\mathcal{N}}
\newcommand{\ham}{\mathcal{H}}

\newcommand{\CKM}{V^{\mathrm{CKM}}}
\newcommand{\CKMd}{V^{\mathrm{CKM} \dagger}}
\newcommand{\BR}{{\mathrm {BR}}}
\newcommand{\Au}{{\mathrm {Au}}}
\newcommand{\Al}{{\mathrm {Al}}}
 
 \newcommand{\MeV}{{\mathrm {MeV}}}
  \newcommand{\GeV}{{\mathrm {GeV}}}
   \newcommand{\TeV}{{\mathrm {TeV}}}

        \newcommand{\NP}{{\mathrm {NP}}}
        \newcommand{\inter}{{\mathrm {int}}}
        \newcommand{\SM}{{\mathrm {SM}}}

\DeclareMathOperator{\sz}{S^1/\mathbb{Z}_2}
\DeclareMathOperator{\szz}{S^1/\mathbb{Z}_2 \times \mathbb{Z}_2^\prime}
\DeclareMathOperator{\z}{\mathbb{Z}_2}

\def\bar#1{\overline{#1}}

\def\inv{^{\raise.15ex\hbox{${\scriptscriptstyle -}$}\kern-.05em 1}}
\def\lbar{{\lower.35ex\hbox{$\mathchar'26$}\mkern-10mu\lambda}} 

\def\to{\rightarrow}

\let\<=\langle
\let\>=\rangle

\let\+=\uparrow

\newcommand{\newc}{\newcommand}
\newc{\gsim}{\lower.7ex\hbox{$\;\stackrel{\textstyle>}{\sim}\;$}}
\newc{\lsim}{\lower.7ex\hbox{$\;\stackrel{\textstyle<}{\sim}\;$}}

\def\OO{\mathcal{O}}

\title{Rare Flavor Processes in Maximally Natural Supersymmetry}

\author[a]{Isabel Garc\'ia Garc\'ia,}
\emailAdd{isabel.garciagarcia@physics.ox.ac.uk}

\author[a,b]{John March-Russell}
\emailAdd{jmr@thphys.ox.ac.uk}

\affiliation[a]{Rudolf Peierls Centre for Theoretical Physics,
University of Oxford,\\
1 Keble Road, Oxford,
OX1 3NP, UK}
\affiliation[b]{Stanford Institute for Theoretical Physics, Department of Physics,\\
Stanford University, Stanford, CA 94305, USA}

\abstract{We study CP-conserving rare flavor violating processes in the recently proposed theory of Maximally Natural Supersymmetry (MNSUSY).
MNSUSY is an unusual supersymmetric (SUSY) extension of the Standard Model (SM) which, remarkably, is un-tuned at present LHC limits.
It employs Scherk-Schwarz breaking of SUSY by boundary conditions upon compactifying an underlying 5-dimensional (5D) theory down to 4D,
and is not well-described by softly-broken $\mathcal{N}=1$ SUSY, with much different phenomenology than the
Minimal Supersymmetric Standard Model (MSSM) and its variants.
The usual CP-conserving SUSY-flavor problem is automatically solved in MNSUSY due to a residual almost exact $U(1)_R$ symmetry,
naturally heavy and highly degenerate 1st- and 2nd-generation sfermions, and heavy gauginos and Higgsinos.
Depending on the exact implementation of MNSUSY there exist important new sources of flavor violation involving
gauge boson Kaluza-Klein (KK) excitations.
The spatial localization properties of the matter multiplets, in particular the brane localization of the 3rd generation states,
imply KK-parity is broken and {\it tree-level}
contributions to flavor changing neutral currents are present in general.
Nevertheless, we show that simple variants of the basic MNSUSY model are safe from present flavor constraints
arising from kaon and $B$-meson oscillations, the rare decays $B_{s,d} \to \mu^+ \mu^-$, $\mu \to {\bar e}ee$ and $\mu$--$e$ conversion in nuclei.
We also briefly discuss some special features of the radiative decays $\mu \to e \gamma$ and ${\bar B}\to X_s \gamma$.
Future experiments, especially those concerned with lepton flavor violation, should see
deviations from SM predictions unless one of the MNSUSY variants with enhanced flavor symmetries is realized.}

\begin{document}

\maketitle


\section{Introduction}
\label{sec:intro}

Any theory that claims to be a UV completion of the Standard Model (SM) implies the introduction of some new mass scale $\Lambda$ potentially much
larger than the electroweak vacuum expectation value (vev), $v$.  The Hierarchy Problem (HP) is the statement that taking radiative corrections into account
the Higgs mass-squared parameter, $m_\phi^2$, is quadratically sensitive to the masses of new particles, expected to be of order $\Lambda$,
so typically driving $m_\phi^2$, and thus $v$, orders of magnitude above its desired value unless a very non-trivial cancellation occurs. 

Softly-broken SUSY is an attractive solution to the HP~\cite{Dimopoulos:1981zb} if the 
symmetry breaking between particle and sparticle masses is not too large.  However, current collider bounds on the sparticle masses imply
that the most popular SUSY theories, those based on the MSSM and its variants, must all be fine-tuned to $\lsim 1\%$,
a level that many physicists find unpalatable \cite{Gherghetta2013,Arvanitaki2013,Hardy2013,Feng2013,Gherghetta2014,Fan2014}.
Recently, a new SUSY solution to the HP, Maximally Natural Supersymmetry (MNSUSY), has been proposed \cite{MNSUSY} that is 
simultaneously
fully natural and consistent with
LHC and other bounds by adopting a radically different way of embedding the SM in a SUSY theory.  In short, MNSUSY is a 4D (excluding the gravitational sector) theory of the weak scale that arises from SUSY in 5 dimensions, with the fifth dimension being compactified on an orbifold line segment with non-local (from the 5D perspective) SUSY-breaking of Scherk-Schwarz \cite{Scherk:1979zr,Scherk:1978ta} type.\footnote{For
related works using Scherk-Schwarz supersymmetry breaking in a similar fashion see Refs.\cite{Barbieri:2003kn,Barbieri:2002sw,Barbieri:2002uk,Barbieri:2000vh}. For earlier related work concerning flavor and other precision constraints
in extra-dimensional models see Ref.\cite{Delgado:1999sv}.}
Scherk-Schwarz supersymmetry breaking (SSSB) is known to be of an extremely soft form \cite{Antoniadis:1998sd,Delgado:1998qr,Kim:2001re}, with all soft SUSY-breaking parameters being {\it finite} and {\it UV insensitive}, in contrast to MSSM-like theories where there are both UV-sensitive
logarithmic enhancements of the soft terms, and connected renormalization group evolution of the soft parameters leading to the gluino mass pulling up the stop mass and therefore increasing tuning (the so-called `gluino-sucks' problem~\cite{Arvanitaki2013}).

A new theory that extends the SM trying to solve the HP necessarily implies the presence of some extra structure, \ie \ new particles,
new interactions and possibly new parameters. As a result, any such theory can affect theoretical predictions
of physical observables with respect to the SM, of particular interest being how this extra structure affects rare flavor violating processes
such as flavor changing neutral currents (FCNC). In the SM, FCNC are absent at tree-level, only arising at one loop and being suppressed 
by the GIM mechanism.
The fact that, to date, there is good agreement between SM predictions and 
experimental measurements of physical observables related
to flavor changing processes sets very strong constraints on how extensions of the SM can contribute to FCNC.
For example, in MSSM-like theories FCNC also arise at one loop, with the new contributions arising from diagrams with SUSY
particles (sfermions, gauginos, Higgsinos) propagating in the loop. As a result, the flavor structure of sfermion mass-squared matrices are
strongly constrained, leading to the SUSY flavor problem that afflicts mediation mechanisms of SUSY-breaking, most notably gravity-mediated
SUSY breaking.

Although in a well-defined sense SSSB is a form of gravity mediation, its unique features imply that the SUSY flavor problem is under much better
control than usual.  Specifically the MNSUSY implementation of SSSB has large and dominantly flavor-universal (at least for the 1st and 2nd generation states)
soft masses.  In addition, there is an approximate $U(1)_R$ symmetry that implies that the dangerous Majorana contributions to the gaugino masses are
small compared to the dominant Dirac gaugino masses, and that $A$-terms are also small, features that are well-known to suppress dangerous SUSY-generated contributions
to FCNC processes \cite{Kribs2008}.   These issues are discussed in Sections~\ref{sec:reviewRsymmetry} and \ref{sec:statusSUSYflavor}.

More importantly, we will be concerned with how in MNSUSY {\it tree-level} FCNC processes may arise, depending on the precise implementation of the MNSUSY
framework, due to the exchange of Kaluza-Klein (KK) excitations of the neutral gauge bosons, and how satisfying the resulting constraints
impacts model building in MNSUSY theories.  We will argue that MNSUSY theories have the feature that new contributions to rare flavor
processes are, given some simple discrete choices, sufficiently small to satisfy current constraints, but typically not too small as to be
completely inaccessible to future experiments.   

Specifically, in Section~\ref{sec:review} we review the basic relevant features of MNSUSY theories while in
Section~\ref{sec:explanationTree} we explain how new tree-level contributions to rare processes can arise.
Regarding processes potentially affected by FCNC at tree-level, in Section~\ref{sec:limits} we study the situation in generalised MNSUSY models (with flavor rotation
matrices left unspecified) for kaon and $B$-meson mixing\footnote{We also make some comments concerning the parameter $\epsilon_K$, a measurement of CP violation in the kaon sector.  We leave a full study of CP-violation
constraints on MNSUSY theories to a future work.} along with the rare decays $B_{s,d} \rightarrow \mu^+ \mu^-$, while in the leptonic sector we consider the processes of $\mu$-$e$ conversion in nuclei and the decay $\mu \rightarrow {\bar e}ee$.
We emphasize that many of our results are more generally applicable to extra-dimensional theories with SM fermions differently localized
and do not depend upon SSSB or the existence of SUSY at all.
In Section~\ref{sec:consequences} we discuss the consequences of the resulting limits  
for the structure of flavor rotation matrices, and thus for the localization properties of the matter multiplets in variant MNSUSY theories, showing that simple variants
are safe from present constraints. We also briefly discuss in Section~\ref{sec:dipole} some special features of the radiative decays $\mu \rightarrow e \gamma$
and $\bar B \rightarrow X_s \gamma$ that apply in MNSUSY theories.
Finally,  Section~\ref{sec:conclusions} contains our conclusions.


\section{Structure of Maximally Natural Supersymmetry}
\label{sec:review}

Here, we will only review the basic aspects of the MNSUSY framework relevant for our study. A more thorough
explanation of the details of the theory can be found in \cite{MNSUSY, moreMNSUSY}.

\subsection{5D SUSY with Scherk-Schwarz breaking}
\label{sec:reviewExtraDim}

For a theory with SM particles in extra dimensions to be a realistic theory of the world, these must be compactified. If the compactification
is a circle $S^1$ of radius $R$, a field propagating in the bulk of the extra dimension is equivalent to an infinite tower of fields from the 4D perspective,
the so-called Kaluza-Klein (KK) modes. Such compactification breaks 5D Lorentz invariance down to 4D, and results
in momentum along the fifth dimension being discretised in multiples of the compactification scale $1/R$. Although 5D Lorentz invariance
is broken, translation along the fifth dimension remains a symmetry of the theory and the now discrete 
momentum along the fifth dimension remains conserved (KK-number conservation).

A true theory of nature, however, must include chiral fermions in the 4D effective theory, and in order to achieve this the compactification cannot
be a circle but needs to be an orbifold, \eg \ $\sz$, the physical length of the extra dimension being $\pi R$.
The fixed points of the orbifold break continuous translation invariance in the fifth dimension,
leading to non-conservation of KK-number, but a translation by $\pi R$ is still a symmetry of the orbifold and a remaining
$\z$ symmetry, known as KK-parity, remains a good symmetry as long as brane-localized interactions are the same at the two
fixed points (this is the case in models like minimal Universal Extra Dimensions).
In the case of MNSUSY, the interactions and matter content on the two fixed point branes are not identical
so KK-parity will be broken, and the theory is more properly described as a $\szz$ orbifold (with $R \rightarrow 2R$).

The minimal ($\N=1$) 5D bulk supersymmetry of MNSUSY corresponds to $\N=2$ SUSY from the 4D perspective.
At $y=0,\pi R$ there are the two inequivalent fixed points of the orbifold that correspond to 4D branes,
each of which can in principle contain 4D $\N=1$ SUSY matter and interactions.
It is useful to recall how the bulk `$\N =2$' SUSY can be written in terms of $\N=1$ superfields.
An $\N=2$ vector supermultiplet can be written in terms of one $\N=1$ vector superfield $V$ and one
$\N=1$ chiral superfield $\Sigma$, both in the adjoint adjoint representation of the gauge group.
On-shell\footnote{$\lambda$ and $\lambda^\prime$ denote two-component Weyl fermions.
Similarly for $\psi$ and $\psi^c$ appearing later.}:
$V^a = (V^a_\mu, \lambda^a)$ and $\Sigma^a = (\sigma^a, \lambda^{\prime a})$.
In addition, an $\N=2$ matter superfield, known as {\it hypermultiplet}, can be written in terms of two $\N=1$ chiral superfields
$\Phi$ and $\Phi^c$, such that $\Phi$ and $\Phi^{c \dagger}$ have the same gauge quantum numbers, with on-shell
degrees of freedom $\Phi = (\phi, \psi)$ and $\Phi^c = (\phi^c, \psi^c)$.

In the phenomenologically interesting range $1/R$ varies from around 4~TeV to 16~TeV, with
associated fine-tuning of $50 \%$ to $3 \%$ (see Figure~4 in \cite{MNSUSY}).
The theory possesses a cutoff, $M_5$, which acts as the fundamental Planck scale of the theory.
We take it to satisfy $M_5 = N/R$ with $N \sim 10$ ($N$ counts the number of KK-modes up to the cutoff),
which is consistent with `naive dimensional analysis' (NDA)
estimates and unitarity bounds on scattering \cite{MNSUSY}.\footnote{Either
extra purely gravitational dimensions, or embedding in a construction such as little string theory can explain
the large observed value of $M_{\rm Planck}$ in terms of the much smaller underlying scale $M_5$ \cite{MNSUSY,moreMNSUSY}.}

\begin{figure}
	\centering
	\includegraphics[scale=0.25]{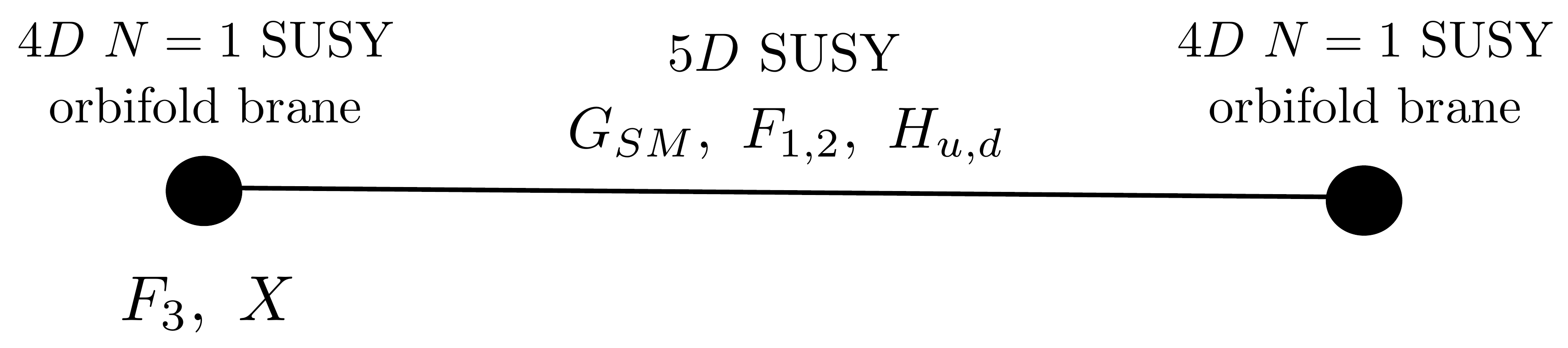}
\caption{Schematic of the basic elements of the MNSUSY model from Ref.\cite{MNSUSY}.  In the 5D bulk are the SM gauge fields, the Higgs
doublets $H_{u,d}$, the two lower generation families $F_{1,2}$, and all the superpartners of these fields as implied by 5D SUSY.  In the minimal MNSUSY model the full 3rd
generation of 4D $\N=1$ chiral multiplets are localized on the brane along with a SM-neutral chiral field $X$.  SUSY is broken non-locally and completely by Scherk-Schwarz boundary conditions, which then causes $F_X\neq 0$. Other choices for the localization of the matter multiplets within $F_{1,2,3}$ are possible while still maintaining low fine-tuning and are motivated by flavor considerations (see Section~\ref{sec:consequences}).} 
\label{fig_location} 
\end{figure}

The structure of MNSUSY, as presented in \cite{MNSUSY}, is such that the gauge and Higgs sectors propagate in the 5D bulk of the extra dimension,
together with the 1st and 2nd family of matter hypermultiplets. The 3rd generation, on the other hand, remains
localized on the $y=0$ brane.
By localizing the 3rd generation of matter in the $y=0$ brane we are introducing an explicit source of KK-parity violation
immediately into the SM-charged sector,
for we are introducing interaction terms in the $y=0$ brane that are not mirrored in the $y=\pi R$ brane.
This structure implies that there is an $\N=2$ vector superfield for each gauge group, two Higgs hypermultiplets
$\mathbb{H}_u=\{H_u, H_u^{c}\}$ and $\mathbb{H}_d=\{H_d, H_d^{c}\}$ and $5$ hypermultiplets
(corresponding to $Q_i$, $\bar U_i$, $\bar D_i$, $L_i$ and $\bar E_i$)
for each of the first two families of matter fields, that we refer to as $\mathbb{F}_{1,2} = (F_{1,2}, F^c_{1,2})$.
The 3rd generation of matter, being brane localized, just consists of the usual $\N=1$ chiral supermultiplets, $F_3$.
This basic setup is illustrated in Figure \ref{fig_location}.\footnote{The Higgs sector of MNSUSY is unusual, with only
a single doublet $H_u$ coupling to matter and acquiring a vev \cite{Davies:2011mp} thus modifying flavor model-building compared
to the MSSM.  Also note that in the version of MNSUSY presented in Ref.\cite{MNSUSY}
the gauge sector of the SM was extended by a $U(1)^\prime$ in order
to raise the physical Higgs mass to its observed value. However, this latter ingredient of the theory is optional, for the
observed Higgs mass may be achieved by other means \cite{moreMNSUSY}, and so we will not take the $U(1)^\prime$ into
account here.}
In addition to this, there is also a chiral superfield $X$ that is a SM-singlet and is localized in the $y=0$ brane and
whose $F$-term picks up a non-zero vev.
It is worth remembering that although two Higgs hypermultiplets are present, only the scalar component of $H_u$ gets
a non-zero vev, and mass terms for down-type quarks and charged leptons are generated
via \kahler terms involving both $H_u$ and the SM-singlet $X$.
As we will see in Section~\ref{sec:explanationTree},
it is this different localization of the three generations of matter along the extra dimension that leads to tree-level FCNC.
A variant of this which is equally un-tuned consists in localizing only the chiral superfields $Q_3$, $\bar U_3$ and $\bar E_3$
on the $y=0$ brane, while the rest of the 3rd generation is allowed to propagate in the bulk. Alternatively, all three families may be
localized on the $y=0$ brane at an EWSB tuning price of $\sim 15\%$. 
As we will show in Section~\ref{sec:consequences}, the former of these variants has reduced
but still potentially observable contributions to rare flavor processes, whereas for the latter these are likely unobservably small.

Whereas a compactification on $\sz$ breaks $\N=2$ SUSY down to 4D $\N=1$,
the remaining $\N=1$ SUSY is broken via the SSSB mechanism \cite{Scherk:1979zr,Scherk:1978ta,Antoniadis:1998sd,Delgado:1998qr}
(equivalent to compactification on an orbifold $\szz$ with $R \rightarrow 2R$),
which corresponds to SUSY breaking by boundary conditions (bc's).
It is the choice of bc's (or, equivalently, of the parities under the two $\z$ symmetries) that fixes
the spectrum of the KK-tower for the different fields.  Specifically, with $\pm$ corresponding to Neumann and Dirichlet respectively, and $(\pm,\pm)$ indicating
bc's satisfied at $y=(0,\pi R)$ branes: $(+,+)$ $m_n=n/R$,  $(-,-)$ $m_n=(n+1)/R$, while $m_n=(2n+1)/(2R)$ for both $(+,-)$ and $(-,+)$, with $n=0,1,\dots$.
Only those fields with bc's $(+,+)$ acquire a massless $0$-mode.
The bc's for the different fields of the theory can be found in Table~\ref{table-bcs} following Ref.\cite{MNSUSY}.
SM particles arising from bulk fields (but {\it not} brane localized states) get a tower of KK-excitations with masses given by
$n/R$, whereas for their 5D $\N=1$ superpartners the tower has masses $(2n+1)/(2R)$.
For a compactification scale $1/R = 4$~TeV, the first KK-excitation of SM particles
sits at 4~TeV, whereas the lowest mode of the supersymmetric partners (gauginos, Higgsinos and sfermions of the
first two generations) are at $1/(2R) = 2$~TeV.
In MNSUSY, Higgsinos get mass without the need of a $\mu$-term, therefore solving the $\mu$-problem
automatically and eliminating the tree-level tuning present in the expressions for the Higgs soft mass-squared.

The brane-localized sfermions first pick up soft masses at 1-loop from gauge interactions with the bulk vector multiplets
(and via the top-Yukawa interaction with the bulk Higgs hypermultiplets for the stop) of the form (\cite{MNSUSY} following \cite{Antoniadis:1998sd})
\begin{equation}
\delta{\tilde m_i}^2 \simeq  \frac{7 \zeta(3)}{16 \pi^4 R^2} \left( \sum_{I=1,2,3} C_I(i) g_I^2 + C_t(i) y_t^2 \right) \equiv \frac{1}{R^2} \delta_i.
\label{eq:1loopsoftmass}
\end{equation}
Here $C_I(i)$, $C_t(i)$ are $\OO(1)$ group theory coefficients given in \cite{MNSUSY}. 

Finally, an additional source of SUSY breaking is the $F$-term of the brane-localized field $X$,
$\langle F_X \rangle \lsim 1/R^2$ \cite{MNSUSY}. 
It will be important that for the bulk superpartners this leads to a parametrically small contribution compared to the direct
SSSB term, see Section~\ref{sec:statusSUSYflavor}.

\begin{table}[h]
  \begin{center}
    \begin{tabular}{l|c|c|c|c}
	5D supermultiplet & $(+,+)$ & $(+,-)$ & $(-,+)$ & $(-,-)$ \\
	\hline
	${\mathbb{V}}^a=\{ V^a, \Sigma^a\}$ & $V^a_\mu$ & $\lambda^a$ & $\lambda^{\prime a} $ & $\sigma^a $ \\
	${\mathbb{H}}_{u,d}=\{H_{u,d}, H_{u,d}^{c}\}$ & $h_{u,d}$ & $\tilde h_{u,d}$ & $\tilde h^c_{u,d}$ & $h^c_{u,d}$ \\
	${\mathbb{F}}_{1,2}=\{F_{1,2}, F_{1,2}^{c}\}$ & $f_{1,2}$ & $\tilde f_{1,2}$ & $\tilde f^c_{1,2}$ & $f^c_{1,2}$ \\
    \end{tabular}
    \caption{Boundary conditions for basic bulk fields of MNSUSY model of Ref.\cite{MNSUSY}. 
	Rows show 5D supermultiplet content. Differing component field bc's non-locally (and fully) break SUSY.  
	$f_{1,2}$ stands for 1st and 2nd generation fermions, $\tilde f_{1,2}$ their 4D $\N=1$ sfermion partners, while
	additional states ($f^c_{1,2}$, $\tilde f^c_{1,2}$) are their 5D SUSY partners.
	Similarly, $h_{u,d}$ refer to the scalar components of $H_{u,d}$ and $\tilde h_{u,d}$ their 4D $\N=1$ superpartners (Higgsinos),
	whereas additional states ($h^c_{u,d}$, $\tilde h^c_{u,d}$) are their 5D SUSY partners.
	Only the $(+,+)$ fields have a massless zero mode.
	These massless zero modes along with the brane localized 3rd family states realise the SM degrees of freedom plus typically some
	additional sub-TeV sfermion states.  At each KK-level states mix to produce mass eigenstates.
\label{table-bcs}} 
  \end{center}
\end{table}

\subsection{R-symmetry structure of MNSUSY}
\label{sec:reviewRsymmetry}

The choice of bc's described above leads to the theory preserving an accidental R-symmetry,
which is exact in the absence of gravitational interactions.
The fermionic components of the $\N=2$ vector supermultiplet ($\lambda$ and $\lambda^\prime$) can partner
to result in Dirac gaugino masses of $1/(2R)$, and similarly for the
two fermionic components of the two Higgs hypermultiplets ($\tilde h_{u,d}$ and $\tilde h^c_{u,d}$),
resulting in Dirac Higgsinos.
Moreover, the R-symmetry is not broken by the vev of the scalar component of $H_u$,
or by Yukawa terms in the superpotential, or by $F_X$. The R-charges of the different fields are given
(in $\N=1$ language) in Table \ref{tab:Rcharges}.
\begin{table}[h]
  \begin{center}
    \begin{tabular}{ l | c | c }
	$\N=1$ superfield & Boson & Fermion \\
	\hline
	$V^a = (V^a_\mu, \lambda^a)$ & $0$ & $+1$ \\
	$\Sigma^a = (\sigma^a, \lambda^{\prime a})$ & $0$ & $-1$ \\
	$H_{u,d} = (h_{u,d}, \tilde h_{u,d})$ & $0$ & $-1$ \\
	$H^c_{u,d} = (h^c_{u,d}, \tilde h^c_{u,d})$ & $+2$ & $+1$ \\
	$F_{1,2,3} = (\tilde f_{1,2,3}, f_{1,2,3})$ & $+1$ & $0$ \\
	$F^c_{1,2} = (\tilde f^c_{1,2}, f^c_{1,2})$ & $+1$ & $0$ \\
	$X = (x, \tilde x)$ & $+2$ & $+1$ 
    \end{tabular}
    \caption{R-charges of relevant fields. The pairs $(\lambda^a, \lambda^{\prime a})$
	and $(\tilde h_{u,d}, \tilde h^c_{u,d})$ have opposite R-charges and partner resulting in
	Dirac gaugino and Higgsino masses. Note that $R(h_u)=0$ and $R(F_X)=0$.}
    \label{tab:Rcharges}  
  \end{center}
\end{table}

This accidental R-symmetry is exact as long as supergravity (SUGRA) interactions are ignored.
At the quantum level, anomaly-mediated contributions to Majorana gaugino masses, sfermion soft masses,
and $A$-terms appear \cite{LutyTASI}.  In terms of the SUGRA conformal compensator $F$-term, $F_\phi$, these are parametrically
(conventionally the scalar component $\phi=1$) \cite{Randall1998}
\begin{equation}
\begin{aligned}
	m_\lambda =& \frac{\beta_g}{g} F_\phi \sim \frac{g^2}{16 \pi^2} F_\phi,~~~
	m_{\tilde f}^2 = -\frac{1}{4} |F_\phi|^2 \frac{d \gamma}{d \log \mu} \sim \frac{g^4}{(16 \pi^2)^2} |F_\phi|^2 \\
	&~{\rm and}~~~~A = \frac{1}{2} \beta_y F_\phi \sim (yg^2~{\rm or}~y^3) \frac{1}{16 \pi^2} F_\phi .
\end{aligned}
\label{eq:AMSUSYbreaking}
\end{equation}

On the other hand, the size of $F_\phi$ is bounded quite generally, including the SSSB case, by \cite{ArkaniHamed2004}
$m^3_{3/2}/(16 \pi^2 M_{\rm Planck}^2) \lesssim F_\phi \lesssim m_{3/2}$
where $m_{3/2}$ is the gravitino mass and $M_{\rm Planck} = \OO(10^{18})$~GeV is the reduced Planck mass.
To be conservative we will assume $F_\phi \sim m_{3/2} \sim 1/(2R)$.\footnote{One must be careful concerning a potential confusion with regard to the gravitational
sector of the theory.  Since the theory has to be embedded in a space-time of dimension $(d+1)> 5$ so as to be able to accommodate the large value of $M_{\rm Planck}$ ($d=6$ is the minimal choice allowed phenomenologically, with two additional spatial dimensions beyond the 5th-dimension already discussed), the full SUGRA algebra is at least $\N=4$ extended from a 4D perspective, and possibly $\N=8$ extended.  As a result there are {\it multiple} gravitini which propagate in two-or-more large spatial dimensions of linear size $L\gg R$ as well
as the $(4+1)$-D space-time already discussed.  A subset of these gravitini pick up a direct Scherk-Schwarz mass of size $1/(2R)\sim 2$ TeV from the R-symmetry-twisted boundary
conditions along the 5th dimension.  However, depending on the UV completion of the model, as well as the embedding of the R-symmetry twist in the full underlying R-symmetry
group, additional gravitini may remain very light with masses parametrically smaller than $1/R$.  Our estimate of $F_\phi$ is maximally pessimistic in that we take
it to be set by the largest possible gravitino mass.  For more discussion of this and other aspects of MNSUSY models we refer the reader to the upcoming Ref.\cite{moreMNSUSY}.}
In this case, we see that the sizes of the anomaly mediated contributions are such that
\begin{equation}
	m_\lambda \sim \frac{g^2}{16 \pi^2} \frac{1}{2R} \ll \frac{1}{2R}~~{\rm and}~~
	m_{\tilde f}^2 \sim \frac{g^4}{(16 \pi^2)^2} \left( \frac{1}{2R} \right)^2 \ll \left( \frac{1}{2R} \right)^2 ,
\end{equation}
\ie \ much smaller than the Dirac gaugino masses and sfermion masses obtained via SSSB.
Similarly, the anomaly mediated $A$-terms result in parametrically small contributions to the sfermion
mass matrix.


\section{Flavor Violation in Maximally Natural Supersymmetry}
\label{sec:FCNCexplanation}

The relevant flavor changing processes that are affected in MNSUSY are, in the quark sector, kaon and $B$-meson oscillations,
which have been observed and measured to high accuracy, and the rare decays $B_{s,d} \rightarrow \mu^+ \mu^-$ and $\bar B \rightarrow X_s \gamma$.
In the lepton sector, we consider three yet-to-be-seen processes for which only experimental upper bounds
exist: $\mu$-$e$ conversion in nuclei and the decays $\mu \rightarrow {\bar e}ee$ and $\mu \rightarrow e \gamma$.
As we show in Section~\ref{sec:explanationTree} these processes can be affected at tree-level in MNSUSY with the exception of the radiative decays $\mu \rightarrow e \gamma$
and $\bar B \rightarrow X_s \gamma$ that only get new contributions from KK-exchange at one loop.  Table~\ref{tab:exp} shows the experimental measurements regarding
these processes and the corresponding SM prediction.

\begin{table}[h]
  \begin{center}
    \begin{tabular}{ l | l | l }
	Observable & Experimental measurement & SM prediction\\
	\hline
	$\Delta M_K$ & $(3.484 \pm 0.006) \cdot 10^{-15}$ GeV \cite{pdg}
				& $(3.1 \pm 1.2) \cdot 10^{-15}$ GeV \cite{SMKaons2011} \\
	$|\epsilon_K|$ & $(2.228 \pm 0.011) \cdot 10^{-3}$ \cite{pdg}
				& $(1.81 \pm 0.28) \cdot 10^{-3}$ \cite{SMKaons2011} \\
	$\Delta M_{B_s}$ & $(1.180 \pm 0.014) \cdot 10^{-11}$ GeV \cite{BMixingLHCb2013}
				& $(1.14 \pm 0.17) \cdot 10^{-11}$ GeV \cite{BsMixingSM2011}\\
	$\Delta M_{B_d}$ & $(3.31 \pm 0.08) \cdot 10^{-13}$ GeV \cite{BMixingLHCb2013}
				& $(3.4 \pm 0.6) \cdot 10^{-13}$ GeV \cite{BsMixingSM2011} \\
	$\BR (B_s \rightarrow \mu^+ \mu^-)$ & $2.9^{+1.1}_{-1.0} \cdot 10^{-9}$ \cite{BmumuLHCb2013}
				& $(3.65 \pm 0.23) \cdot 10^{-9}$ \cite{BsmumuSM2013}\\
	$\BR (B_d \rightarrow \mu^+ \mu^-)$ & $< 7.4 \cdot 10^{-10}$ \cite{BmumuLHCb2013}
				& $(1.07 \pm 0.10) \cdot 10^{-10}$ \cite{BdmumuSM2012}\\
	$\BR (\mu \Au \rightarrow e \Au)$ & $< 7 \cdot 10^{-13}$ \cite{SINDRUMmue2006}
				& $\lesssim 10^{-50}$ \\
	$\BR (\mu \rightarrow {\bar e} e e)$ & $< 1.0 \cdot 10^{-12}$ \cite{SINDRUMmu3e1988}
				& $\lesssim 10^{-50}$ \\
	$\BR (\mu \rightarrow e \gamma)$ & $< 5.7 \cdot 10^{-13}$ \cite{MEG2013}
				& $\lesssim 10^{-50}$ \\
	$\BR (\bar B \rightarrow X_s \gamma)$ & $ (3.55 \pm 0.24) \cdot 10^{-4}$ \cite{HFAG2012}
				& $ (3.15 \pm 0.23) \cdot 10^{-4}$ \cite{Misiak2010a}
    \end{tabular}
    \caption{Experimental measurements together with their corresponding SM predictions.
	Similar experimental results regarding $\BR (B_s \rightarrow \mu^+ \mu^-)$ and $\BR (B_d \rightarrow \mu^+ \mu^-)$
	can be found in \cite{CMSBmesonDecays}.}
    \label{tab:exp}  
  \end{center}
\end{table}

Apart from a limited discussion of $\epsilon_K$ we do not here consider CP-violating observables.  In this work we
assume that {\it any potential new phases are either zero or tiny}. We hope to return to the
physics of CP-violation in MNSUSY in a subsequent publication.

\subsection{Status of the SUSY flavor problem in MNSUSY}
\label{sec:statusSUSYflavor}

In SUSY extensions of the SM consideration of potentially flavor-violating loop processes involving squarks and sleptons
(and gauginos/Higgsinos) traditionally leads to strong constraints on the form of the sfermion mass-squared matrices.  We now explain
why the situation in MNSUSY theories is generically much improved.

First, as noted in \cite{Kribs2008}, supersymmetric theories with an accidental R-symmetry larger than $\z$ present some very attractive
features. For example, $A$-terms are forbidden, and therefore do not contribute to flavor changing processes, and the Dirac nature of gaugino
and Higgsino masses leads to the elimination of an important set of 1-loop flavor-changing diagrams.
Similarly, $\Delta L=1$ and $\Delta B=1$ operators typically present in MSSM-like theories are also forbidden by the
R-symmetry, as well as dimension-5 operators leading to proton decay. As we have explained in Section~\ref{sec:reviewRsymmetry},  
in MNSUSY theories there automatically exists a residual R-symmetry that is only broken by tiny loop-level effects.  This feature greatly
suppresses the traditional 1-loop flavor violation that occurs in MSSM-like theories.

The leading source of off-diagonal sfermion mass terms allowed by the R-symmetry are the
brane-localized \kahler terms involving $F_X$.  Importantly, such terms are parametrically small
compared to the dominant sfermion mass-squareds arising from SSSB, at least for those states in the bulk. 
Consider e.g. left-handed (LH) squarks with relevant 4D Lagrangian terms
\begin{equation}
\begin{aligned}
	& \int d^4 \theta dy \delta(y) \frac{X^\dagger X}{M_5^2} \left(	
			Q_3^\dagger c^Q_{33} Q_3 + \frac{1}{M_5} \sum_{i,j=1}^2 Q_i^{5D \dagger} c^Q_{ij} Q_j^{5D}
			+ \frac{1}{\sqrt M_5} \sum_{i=1}^2 \left( Q_i^{5D \dagger} c^Q_{i3} Q_3 + {\rm h.c} \right) \right)\\
				&= \frac{|F_X|^2}{M_5^2} \left(	
			\tilde Q_3^\dagger c^Q_{33} \tilde Q_3 + \frac{1}{M_5} \sum_{i,j=1}^2 \tilde Q_i^{5D \dagger} c^Q_{ij} \tilde Q_j^{5D}
			+ \frac{1}{\sqrt M_5} \sum_{i=1}^2 \left( \tilde Q_i^{5D \dagger} c^Q_{i3} \tilde Q_3 + {\rm h.c} \right) \right) + \dots
\end{aligned}
\label{eq:mass2squarks}
\end{equation}
where $c^Q$ is a general Hermitian $3\times 3$ matrix.
4D-normalizing the states, including the contribution from SSSB (both direct and the 1-loop radiative term, $\delta$, for the 3rd family given in Eq.(\ref{eq:1loopsoftmass})),
the LH squarks 0-mode mass-squared matrix is
\begin{equation}
	 \begin{pmatrix} \tilde u^{(0) \dagger}_{L}, & \tilde c^{(0) \dagger}_{L}, & \tilde t_L^\dagger \end{pmatrix} \frac{1}{R^2}
				\left\{ \frac{1}{4} {\rm diag}(1,1,0) + \Delta^Q \right\}
	\begin{pmatrix} \tilde u^{(0)}_{L} \\ \tilde c^{(0)}_{L} \\ \tilde t_L\end{pmatrix}
				+ (\{ u , c, t\} \rightarrow \{ d ,s, b\})
\label{eq:mass2squarksexplicit}
\end{equation}
with (using $|\langle F_X \rangle |^2 / M_5^2 \lsim (R^2 M_5)^{-2} = (N R)^{-2}$, and $N= M_5 R \sim 10$)
\begin{equation}
	\Delta^Q = 
	\begin{pmatrix} \frac{c^Q_{11}}{\pi N^3}& \frac{c^Q_{12}}{\pi N^3} & \frac{c^Q_{13}}{\pi^{1/2}N^{5/2}}  \\
					\frac{c^Q_{21}}{\pi N^3}& \frac{c^Q_{22}}{\pi N^3} &  \frac{c^Q_{23}}{\pi^{1/2}N^{5/2}} \\
					\frac{c^Q_{31} }{\pi^{1/2}N^{5/2}}& \frac{c^Q_{32} }{\pi^{1/2}N^{5/2}} & \delta+\frac{c^Q_{33}}{N^2} \end{pmatrix} .
\label{eq:DeltaQ}
\end{equation}
Similar contributions are present for right-handed (RH) squarks and for sleptons.
The analogous matrices for other fields will be denoted by $c^\Phi$ and $\Delta^\Phi$ with
$\Phi=\bar D, \bar U, L, \bar E$.
For loops involving sfermions and gauginos/Higgsinos we work
in the basis where fermion-sfermion-gaugino/Higgsino interactions are flavor diagonal, treating off-diagonal entries in
sfermion mass-squared matrices in the mass-insertion approximation.

\subsection{FCNC at tree-level from KK modes}
\label{sec:explanationTree}

The presence of tree-level FCNC in the basic version of MNSUSY
has its root in the fact that the 3rd generation is localized at $y=0$
whereas the lower generations propagate in the 5D bulk, along with the gauge and Higgs sectors.
The 0-modes of neutral gauge bosons ($Z$, photon $\gamma$ and gluons $g^a$)
couple to fermions in a flavor-preserving way, as it must be, since from the perspective of the 4D effective theory
the SM structure must be recovered.  But as we now show in detail higher KK-modes couple in a way that is not generation universal, therefore introducing
a new source of flavor violation beyond the Yukawa couplings to the Higgs.  Moreover these interactions are, in general, KK-parity violating.
These two features imply that the theory contains new sources of {\it tree-level} FCNC.

To illustrate this, it is useful to consider the 5D action and see how the 4D Lagrangian arises after we integrate
over the fifth dimension. For pedagogical simplicity we focus upon the photon field and its interactions with
down-type quarks (the interactions of the other neutral bosons with matter similarly follow). The relevant terms in the
4D Lagrangian are\footnote{For convenience, we use 4-component Dirac fermion notation from now on.}
\begin{equation}
 \lang^\gamma_4 = \int_0^{\pi R}\!\!\!\!\!\! dy \ e_{5D} Q_d A_\mu^{5D} (x,y)
		\biggl\{ \sum_{q=d,s} \bar q^{5D} (x,y) \gamma^\mu q^{5D} (x,y)
		+ \delta(y) \ \bar b(x) \gamma^\mu b(x) \biggr\}
				 \equiv \lang^{\gamma (0)}_4 + \lang^{\gamma ({\rm KK})}_4 + \dots
\label{eq:5Daction}
\end{equation}
where $e_{5D}$ refers to the electromagnetic coupling in its 5D normalization, so $e_{5D} = e \sqrt{\pi R} > 0$, $Q_d = -1/3$,
and we have chosen to write
$\lang^{\gamma}_4$ as a sum of two terms $\lang^{\gamma (0)}_4$ and $\lang^{\gamma ({\rm KK})}_4$ that contain the interactions with only quark 0-modes
and with one quark 0-mode and one higher KK-mode respectively. The dots represent terms containing two quark KK-modes
that we will not make explicit here. The 5D fields $A_\mu^{5D}$ and $q^{5D}$ (both with
positive parity under both $\z$ symmetries) can be written as
\begin{equation}
\begin{aligned}
	q^{5D} (x,y) &= \sum_{n=0}^{\infty} q^{5D (n)} (x) \cos \frac{n y}{R} 
				= \frac{1}{\sqrt{\pi R}} q^{(0)}(x) + \sqrt{\frac{2}{\pi R}} \sum_{n=1}^{\infty} q^{(n)} (x) \cos \frac{n y}{R} \\
	A_\mu^{5D} (x,y) &= \sum_{n=0}^{\infty} A_\mu^{5D (n)} (x) \cos \frac{n y}{R} 
				= \frac{1}{\sqrt{\pi R}} A_\mu^{(0)}(x) + \sqrt{\frac{2}{\pi R}} \sum_{n=1}^{\infty} A_\mu^{(n)} (x) \cos \frac{n y}{R}
\end{aligned}
\end{equation}
where $\phi^{(n)}$ corresponds to the KK-modes of the corresponding field in their appropriate 4D normalization and the 0-modes
$q^{(0)}$ and $A_\mu^{(0)}$ are to be identified with the SM quarks and photon respectively.
Retaining only those terms involving the 0-modes of the quark fields,
\begin{equation}
\begin{aligned}
\lang^{\gamma (0)}_4 	&= \int_0^{\pi R}\!\!\!\!\!\! dy \ \frac{e_{5D}}{\sqrt{\pi R}} Q_d
			\left( A_\mu^{(0)} + \sqrt2 \sum_{n=1}^{\infty} A_\mu^{(n)} \cos \frac{n y}{R} \right)
			\left( \frac{1}{\pi R} \sum_{q=d,s} \bar q^{(0)} \gamma^\mu q^{(0)} + \delta(y) \bar b \gamma^\mu b \right)\\
		&= e Q_d A_\mu^{(0)} (\bar d^{(0)} \gamma^\mu d^{(0)} +\bar s^{(0)} \gamma^\mu s^{(0)} +\bar b \gamma^\mu b)
			+ \sqrt2 e Q_d \left( \sum_{n=1}^{\infty} A_\mu^{(n)} \right) \bar b \gamma^\mu b\\
		& \equiv e Q_d A_\mu \bar q^d \gamma^\mu q^d
			+ \sqrt2 e Q_d \left( \sum_{n=1}^{\infty} A_\mu^{(n)} \right) \bar q^d \gamma^\mu A q^d
\end{aligned}
\label{eq:5Daction0}
\end{equation}
where we have defined $A_\mu \equiv A_\mu^{(0)}$ and $q^d \equiv ( d,  s,  b )^T$,
with $d \equiv d^{(0)}$ and $s \equiv s^{(0)}$ (and suppressed the $x$-dependence of the KK-fields).
The matrix $A$ encodes the flavor structure of the photon non-zero KK-mode interactions and is (if only the 3rd generation
brane localized)
\begin{equation}
	A = \begin{pmatrix} 0 & 0 & 0 \\ 0 & 0 & 0 \\ 0 & 0 & 1 \end{pmatrix} .
\label{eq:Amatrix3rdgen}
\end{equation}
Eqs.(\ref{eq:5Daction0}) and (\ref{eq:Amatrix3rdgen}) show that the 0-mode of the neutral gauge boson, the SM photon in this example,
couples universally to all three generations, whereas its KK-modes do not. The same conclusion holds for the interactions
of all non-zero gauge boson KK-excitations.
In particular, in this gauge-eigenbasis the non-zero KK-modes
of gauge bosons couple to fermion fields in a way that is flavor diagonal but not flavor universal.
In addition, the generation-non-universal interaction is not proportional to the Yukawa couplings of the matter
fermions, so this flavor violation is not of the Minimal Flavor Violation (MFV) \cite{MFV} type.

When considering variants of the basic MNSUSY setup, as we do in Section~\ref{sec:consequences}, it is important to realise that
there is not just one $A$-matrix, but in principle instead {\it five different $A$-matrices}, $A^\Phi$, each one of which encodes the localization
properties in generation space of the basic matter multiplets of the SM, $\Phi = Q, \bar U, \bar D, L, \bar E$.  If, for example, all
matter were brane localized except for the three generations of ${\bar F}_i= \{ \bar D_i, L_i\}$ (using SU(5) notation) all of which
were taken to propagate in the bulk, then in this `ten-five split' case
\begin{equation}
	A^{(Q,\bar U,\bar E)} = \begin{pmatrix} 1 & 0 & 0 \\ 0 & 1 & 0 \\ 0 & 0 & 1 \end{pmatrix},~~{\rm and}~~A^{(\bar D,L)} = \begin{pmatrix} 0 & 0 & 0 \\ 0 & 0 & 0 \\ 0 & 0 & 0 \end{pmatrix} ,
\label{eq:Amatrix510split}
\end{equation}
and in this variant there is {\it no flavor violation due to KK-gauge-boson exchange at all}.

To work with the physical mass-eigenstates of the matter fermions requires rotating
the LH and RH matter fields by $3 \times 3$ unitary matrices.
For example, for down-type quarks $q^d_X \equiv ( d_X, s_X, b_X )^T \rightarrow R^d_X q^d_X$ for $X=L,R$.
Under these rotations, the interaction between the photon 0-mode and the quark fields remains unchanged, but
the coupling with higher KK-modes is modified $\bar q^d \gamma^\mu A q^d \rightarrow \bar q^d_L \gamma^\mu (R^{d \dagger}_L A R^d_L) q^d_L
+ \bar q^d_R \gamma^\mu (R^{d \dagger}_R A R^d_R) q^d_R$.
Therefore, defining
\begin{equation}
	B^d_L \equiv R^{d \dagger}_L A^Q R^d_L , \qquad B^d_R \equiv R^{d \dagger}_R A^{\bar D} R^d_R ,
\label{eq:mixingMatrix}
\end{equation}
the couplings between the photon KK-modes and mass-eigenbasis down-type quarks are
\begin{equation}
\lang^{\gamma (0)}_4 = e Q_d A_\mu \bar q^d \gamma^\mu q^d
			+ \sqrt2 e Q_d \left( \sum_{n=1}^{\infty} A_\mu^{(n)} \right) (\bar q^d_L \gamma^\mu B^d_L q^d_L
															+ \bar q^d_R \gamma^\mu B^d_R q^d_R)
\label{eq:finalLag}
\end{equation}
Eq.(\ref{eq:finalLag}) shows that the coupling between gauge boson KK-modes and the 0-modes of the matter fields is non-diagonal in the mass-eigenbasis.
Second, it is worth noting that although the SM photon couples to LH and RH fields in exactly the same way, this is not true of its higher modes.
Since the flavor changing interactions we are describing here arise once we rotate from the gauge-eigenbasis to the mass-eigenbasis,
different rotation matrices for LH and RH quarks result in different couplings.

Although we have illustrated the appearance of tree-level FCNC in MNSUSY with the simple example of the photon field and its
couplings to down-type quarks, the same occurs with all the other neutral gauge bosons ($Z$ and gluons $g^a$)
and fermions (up-type quarks and leptons). Since we will consider processes involving quarks and charged leptons,
we will be concerned with the mixing matrices $B^u_{L,R}$, $B^d_{L,R}$ and $B^e_{L,R}$, which are built from the five $A^{(Q,{\bar U},{\bar D},L,{\bar E})}$ matrices
and the rotation matrices $R^u_{L,R}$, $R^d_{L,R}$ and $R^e_{L,R}$.\footnote{If flavor-violating interactions involving LH neutrinos
(and possibly RH $N$'s) were to be of interest then
potentially one more $A^N$ and two more $B^\nu_{L,R}$ matrices are relevant. MNSUSY models present interesting new possibilities
for neutrino physics \cite{ArkaniHamed:1998vp,Hebecker:2002re,Hambye:2004jf} but we postpone this topic to a future publication.}


\section{Evaluation of Constraints}
\label{sec:limits}


As we have explained, tree-level FCNC in MNSUSY are mediated by KK-modes of neutral gauge bosons, with
masses $m_n^2 \approx (n/R)^2$ for $n=1,2...$, and $M\equiv 1/R \gsim 4$ TeV.
As the energy scales for the considered processes are much smaller, it makes sense to use an effective
field theory approach in which the heavy bosons are integrated out giving 4-fermion effective operators that describe the
interaction at low energies. All the IR physics is then encoded in the matrix elements of the effective operators,
whereas effects from scale $1/R$ are included in the Wilson coefficients of the
corresponding operators and their renormalization group (RG) evolution.
Schematically, then, the effective Hamiltonian for a particular process can be written as
$\ham_{eff} = \sum_I C_I (M, \mu) Q_I(\mu)$, where $Q_I$ are the four-fermion effective operators and
$C_I$ the corresponding Wilson coefficients, both at RG scale $\mu$.
The Wilson coefficients need to be RG evolved to the IR scale at
which the matrix elements of the effective operators are evaluated.
In our work we have taken RG effects into account for hadronic processes, but not for processes involving leptons only as they are numerically unimportant.

\subsection{Kaon oscillations}
\label{sec:limitsKMesons}

Whereas the experimental measurement of the mass difference, $\Delta M_K$, in the kaon system is extremely precise
(see Table \ref{tab:exp}), the uncertainty in the SM prediction is $\OO(40\%)$ and so the size of NP contributions should be bounded
by this uncertainty.  For the CP violation parameter $\epsilon_K$ the discrepancy between the SM prediction and the measurement is larger
than the dominant (theoretical) uncertainty, with the prediction being below the measured value, so in principle this leads to a preference
for a NP contribution to increase $|\epsilon_K|$. To be conservative, we set limits by bounding $|\epsilon^{\rm NP}_K|$ by the
$\OO(15\%)$ SM uncertainty.

Given the relevant effective NP Hamiltonian the difference in mass between the two physical states is (see, \eg \ \cite{Blanke2008})
\begin{equation}
	\Delta M_K = 2 {\bf Re} (M^K_{12}) \quad {\rm where} \quad M^K_{12} = \frac{1}{2 M_K} \langle \bar K^0 | \ham_{eff} | K^0 \rangle^*
\label{eq:kaonmassformula}
\end{equation}
whereas $|\epsilon_K|$ is given by
\begin{equation}
	|\epsilon_K| = \frac{\kappa_\epsilon}{\sqrt{2} (\Delta M_K)_{\rm exp}} |{\bf Im} (M^K_{12})|
\end{equation}
with $\kappa_\epsilon=0.92 \pm 0.02$.  The matrix element $\langle \bar K^0 | \ham_{eff} | K^0 \rangle$ can be written as
\begin{equation}
	\langle \ham_{eff} \rangle
	\equiv \langle \bar K^0 | \ham_{eff} | K^0 \rangle
	= \sum_I C_I (M, \mu_0) \langle Q_I(\mu_0) \rangle
\end{equation}
where $\mu_0 = 2$ GeV is the RG scale used in lattice determinations of the matrix elements.

The operator basis relevant for kaon mixing is (here $i,j$ are color indices)
\begin{equation}
\begin{aligned}
	Q_1^K &= (\bar s_L^i \gamma^\mu d_L^i) (\bar s_L^j \gamma_\mu d_L^j) &
		\qquad \tilde Q_1^K &= Q_1^K (L \leftrightarrow R)\\
	Q_2^K &= (\bar s_R^i d_L^i) (\bar s_R^j d_L^j)&
		\qquad \tilde Q_2^K &= Q_2^K (L \leftrightarrow R)\\
	Q_3^K &= (\bar s_R^i d_L^j) (\bar s_R^j d_L^i)&
		\qquad \tilde Q_3^K &= Q_3^K (L \leftrightarrow R)\\
	Q_4^K &=(\bar s_L^i d_R^i) (\bar s_R^j d_L^j)&\qquad Q_5^K &=(\bar s_L^i d_R^j) (\bar s_R^j d_L^i)
\end{aligned}
\label{eq:effK}
\end{equation}
Unlike the SM, in MNSUSY there are tree-level contributions
mediated by neutral gauge boson KK-modes.
In the basis of Eq.(\ref{eq:effK}) we find the non-zero Wilson coefficients at scale $M= \OO(1/R)$ to be
(with $\alpha^f_L=T_{3L}^{f} - Q^f s^2_W$, $\alpha^f_R=- Q^f s^2_W$ as usual, where $s^2_W = \sin^2 \theta_W$)
\begin{equation}
\begin{aligned}
	C_1^K (M) & \approx \frac{\pi^2}{6} \frac{1}{(1/R)^2} (B^d_{L21})^2
		\left( \frac{1}{3} g_s^2 + \frac{e^2}{9} + \frac{g^2}{c_W^2} (\alpha^d_L)^2 \right)\\
	\tilde C_1^K (M) & \approx \frac{\pi^2}{6} \frac{1}{(1/R)^2} (B^d_{R21})^2
		\left( \frac{1}{3} g_s^2 + \frac{e^2}{9} + \frac{g^2}{c_W^2} (\alpha^d_R)^2 \right)\\
	C_4^K (M) &\approx\frac{\pi^2}{6} \frac{1}{(1/R)^2} (B^d_{L21} B^d_{R21}) (-2 g_s^2)\\
	C_5^K (M) &\approx\frac{\pi^2}{6} \frac{1}{(1/R)^2} (B^d_{L21} B^d_{R21})
		\left( \frac{2}{3} g_s^2 - \frac{4 e^2}{9}  - \frac{4 g^2}{c_W^2} \alpha^d_L \alpha^d_R\right).
\end{aligned}
\label{eq:KMesonsWilsonCoeff}
\end{equation}

Using these values as bc's we have numerically computed the coefficients at the IR scale $\mu_0 = 2$ GeV
using the appropriate RG evolution equations \cite{RGKaons1998}. The operators $Q^K_2$, $Q^K_3$ and $\tilde Q^K_2$, $\tilde Q^K_3$
of Eq.(\ref{eq:effK}) are not present at scale $1/R$ and they are not generated by RG evolution.
The hadronic matrix elements can be written as 
\begin{equation}
\begin{aligned}
	&\langle \bar K^0|Q^K_1 (\mu_0) |K^0 \rangle
		= \langle \bar K^0|\tilde Q^K_1 (\mu_0) |K^0\rangle
		= \frac{2}{3} m_K^2 f_K^2 B^K_1 (\mu_0) \\
	&\langle \bar K^0|Q^K_4 (\mu_0) |K^0 \rangle
		= \frac{1}{2} \left\{ \frac{m_K}{m_s(\mu_0) + m_d(\mu_0)} \right\}^2 m_K^2 f_K^2 B^K_4 (\mu_0)\\
	&\langle \bar K^0|Q^K_5 (\mu_0) |K^0 \rangle
		= \frac{1}{6} \left\{ \frac{m_K}{m_s(\mu_0) + m_d(\mu_0)} \right\}^2 m_K^2 f_K^2 B^K_5 (\mu_0)
\end{aligned}
\end{equation}
where $f_K = 157.5 \pm 3.3 \ \MeV$ \cite{KaonsDecayConstant2008} and
$B^K_1 = 0.563 \pm 0.047$, $B^K_4 = 0.938 \pm 0.048$, and $B^K_5 = 0.616  \pm 0.059$ \cite{matrixElementsKaons2006}.
We thus find the MNSUSY contributions to $\Delta M_K$ and $|\epsilon_K|$ are (ignoring the slow dependence of
$\alpha_s$ on $1/R$) approximated by
\begin{equation}
	\frac{\Delta M_K^{\rm NP}}{\GeV} \approx 10^{-10} \left( \frac{4 \ \TeV}{1/R} \right)^2 \{
		-830 {\bf Re} (B^d_{L21} B^d_{R21}) + 1.5 {\bf Re} (B^{d}_{L21})^2 + 1.2 {\bf Re} (B^{d}_{R21})^2 \}
\label{eq:DeltaMKformula}
\end{equation}
\begin{equation}
	|\epsilon_K^{\rm NP}| \approx 10^4 \left( \frac{4 \ \TeV}{1/R} \right)^2
		| 780 {\bf Im} (B^d_{L21} B^d_{R21}) - 1.4 {\bf Im} (B^{d}_{L21})^2 - 1.1 {\bf Im} (B^{d}_{R21})^2 |.
\label{eq:epsilonKformula}
\end{equation}

\subsection{B-meson oscillations}
\label{sec:limitsBMesons}

The mass differences between physical states in the neutral $B$-meson system are given by expressions
analogous to Eq.(\ref{eq:kaonmassformula}), while the basis of effective 4-fermion operators relevant for
$B_s$-meson mixing is
\begin{equation}
\begin{aligned}
	Q_1^s &= (\bar b_L^i \gamma^\mu s_L^i) (\bar b_L^j \gamma_\mu s_L^j)&
		\qquad \tilde Q_1^s &= Q_1^s (L \leftrightarrow R)\\
	Q_2^s &= (\bar b_R^i s_L^i) (\bar b_R^j s_L^j)&
		\qquad \tilde Q_2^s &= Q_2^s (L \leftrightarrow R)\\
	Q_3^s &= (\bar b_R^i s_L^j) (\bar b_R^j s_L^i)&
		\qquad \tilde Q_3^s &= Q_3^s (L \leftrightarrow R)\\
	Q_4^s &=(\bar b_L^i s_R^i) (\bar b_R^j s_L^j)&
		\qquad Q_5^s &=(\bar b_L^i s_R^j) (\bar b_R^j s_L^i) .
\end{aligned}
\label{eq:effB}
\end{equation}
The substitution $s \rightarrow d$ gives those for $B_d$-meson mixing.
The high scale Wilson coefficients are the same as those in Eq.(\ref{eq:KMesonsWilsonCoeff})
after the substitution $B^d_{X21} \rightarrow B^d_{X32}$ and $B^d_{X21} \rightarrow B^d_{X31}$ ($X=L,R$)
for $B_s - \bar B_s$ and $B_d - \bar B_d$ mixing respectively,
while the relevant matrix-element parameters, evaluated via non-perturbative lattice calculations at a scale $\mu_b = 4.6$ GeV, 
and decay constants are given in Refs. \cite{matrixElementsBs2001} and \cite{BMesonsDecayConstants2014}.
Upon numerically computing the Wilson coefficients at the IR scale $\mu_b$
using the RG evolution equations \cite{RGBs2001} we find that the final
MNSUSY contributions to $\Delta M_{s,d}$ are approximated by
\begin{equation}
	\frac{|\Delta M_s^{\rm NP}|}{\GeV} \approx 10^{-9} \left( \frac{4 \ \TeV}{1/R} \right)^2 \left|
		180 (B^{d}_{L32} B^{d}_{R32}) - 6.0 (B^{d}_{L32})^2 - 4.7 (B^{d}_{R32})^2 \right|
\label{eq:DeltaMBsformula}
\end{equation}
\begin{equation}
	\frac{|\Delta M_d^{\rm NP}|}{\GeV} \approx 10^{-9} \left( \frac{4 \ \TeV}{1/R} \right)^2 \left|
		120 (B^{d}_{L31} B^{d}_{R31}) - 4.1 (B^{d}_{L31})^2 - 3.2 (B^{d}_{R31})^2 \right| .
\label{eq:DeltaMBdformula}
\end{equation}

\subsection{Muon decay $\mu \rightarrow {\bar e}ee$}
\label{sec:limitsmu3e}

The upper limit on $\BR(\mu \rightarrow {\bar e}ee)$ of $1.0\cdot 10^{-12}$ is almost
40 orders of magnitude above the SM expectation.
Therefore, any experimental observation of this rare process would be a clear sign of NP, and
in this respect we note the Mu3e experiment will start taking data in 2015 with projected sensitivity
of $10^{-16}$ \cite{mu3eProposal2013}.

In MNSUSY tree-level diagrams can contribute to this rare process, mediated by KK-modes of neutral colorless
gauge bosons. Approximating the total decay rate of the muon by
$\Gamma_\mu^{\rm total} \approx \Gamma_{\mu \rightarrow \nu_\mu e \bar \nu_e}$,
the MNSUSY contribution to this BR is given by
\begin{equation}
	\BR(\mu \rightarrow {\bar e}ee) \approx \frac{1}{2 G_F^2} \left( \frac{\pi^2}{6} \right)^2 \frac{1}{(1/R)^4}
		(2 S_{LL}^e |B^e_{L11} B^e_{L12}|^2 + S_{LR}^e |B^e_{L11} B^e_{R12}|^2 + (L \leftrightarrow R))
\label{eq:BRmu3e}
\end{equation}
where $S_{LL}^e = (g^2/c^2_W) (\alpha^e_L)^2 + e^2$, $S_{LR}^e = (g^2/c^2_W) \alpha^e_L \alpha^e_R + e^2$,
and $S_{RR}^e = S_{LL}^e$, $S_{RL}^e = S_{LR}^e$ with $L \leftrightarrow R$. 
The approximate expression is then
\begin{equation}
\begin{aligned}
	\BR(\mu \rightarrow {\bar e}ee) \approx 3.9 \cdot 10^{-6} \left( \frac{4 \ \TeV}{1/R} \right)^4 \{
		&2.5 |B^e_{L11} B^e_{L12}|^2 + 2.3 |B^e_{R11} B^e_{R12}|^2 +\\
		&0.61 |B^e_{L11} B^e_{R12}|^2 + 0.61 |B^e_{R11} B^e_{L12}|^2 \} .
\end{aligned}
\label{eq:mu3eformula}
\end{equation}

\subsection{$\mu - e$ conversion in nuclei}
\label{sec:mue}

The SM prediction for $\BR(\mu N \to e N)$ is far below the current experimental upper bound of $7\cdot 10^{-13}$ set using gold targets,
so this is a clear test of NP.  The planned aluminium target experiment Mu2e operating at Fermilab has projected BR sensitivity 
$\sim 5.7 \cdot 10^{-17}$ \cite{mu2e2008}.

In MNSUSY the basis of relevant effective operators, for $q=u,d$, is 
\begin{equation}
\begin{aligned}
	Q^q_{VV} &= (\bar e \gamma^\mu \mu)(\bar q \gamma_\mu q)&
			\qquad Q^q_{AA} &= (\bar e \gamma^\mu \gamma^5 \mu)(\bar q \gamma_\mu \gamma^5 q)\\
	Q^q_{VA} &= (\bar e \gamma^\mu \mu)(\bar q \gamma_\mu \gamma^5 q)&
			\qquad Q^q_{AV} &= (\bar e \gamma^\mu \gamma^5 \mu)(\bar q \gamma_\mu q) .
\end{aligned}
\label{eq:effmue}
\end{equation}
However, the hadronic axial current contribution can be neglected compared to the vector current as,
at very low momentum transfer
$\langle N | \bar u \gamma^0 u | N \rangle  \sim 2Z  + N \sim 10^2$
and $\langle N | \bar d \gamma^0 d | N \rangle  \sim Z + 2N \sim 10^2$, 
whereas $\langle N | \bar q \gamma^i q | N \rangle \sim 0$ for $i=1,2,3$ \cite{Kitano2002} and $\langle N | \bar q \gamma^\mu \gamma^5 q | N \rangle\sim S_{nuc} \sim 1$.
Here $Z$ and $N$ refer to the relevant nuclear species, and $S_{nuc}$ its nuclear spin. 

KK-modes of neutral colorless gauge bosons can give tree-level contributions to this process,
and adapting the analysis of Ref.\cite{Langacker2000} we find the approximate formula\footnote{Here
$\Gamma^{\rm cap}_{\mu N}$ is the muon atomic capture
rate ($\Gamma^{\rm cap}_{\mu \Au} \simeq 13 \cdot 10^6 \ {\rm s}^{-1}$,
$\Gamma^{\rm cap}_{\mu \Al} \simeq 0.71 \cdot 10^6 \ {\rm s}^{-1}$ \cite{Kitano2002});
$Z_{eff}^{\Au} = 33.64$, $Z_{eff}^{\Al} = 11.48$ \cite{Suzuki1987} is the effective
nuclear charge in the $1s$ muon atomic state;
$F_p$ is the nuclear form factor at $|\vec p| = m_\mu$
($|F_p^{\Au}| \approx 0.16$, $|F_p^{\Al}| \approx 0.64$ \cite{Kitano2002}).
Finally, $S_{LL}^{de} = (g^2/c^2_W) \alpha^d_L \alpha^e_L + e^2/3$, and $S_{LR}^{de} = (g^2/c^2_W) \alpha^d_L \alpha^e_R + e^2/3$,
while $S_{RR}^{de}$ and $S_{RL}^{de}$ follow with $L \leftrightarrow R$, and similarly for $S^{ue}_{XY}$.}
\begin{equation}
\begin{aligned}
	\BR_{\mu-e} & \approx \left( \frac{\pi^2}{6} \right)^2 \frac{1}{(1/R)^4} \frac{\alpha^3 m_\mu^5}{16 \pi^2 \Gamma^{\rm cap}_{\mu N}}
		\frac{Z_{eff}^4}{Z} |F_p|^2\\
		\{ & |B^e_{L12}|^2 |(2Z+N)(B^u_{L11} S^{ue}_{LL} + B^u_{R11} S^{ue}_{RL}) + (Z+2N)(B^d_{L11} S^{de}_{LL} + B^d_{R11} S^{de}_{RL})|^2+\\
		+ & |B^e_{R12}|^2 |(2Z+N)(B^u_{L11} S^{ue}_{LR} + B^u_{R11} S^{ue}_{RR}) + (Z+2N)(B^d_{L11} S^{de}_{LR} + B^d_{R11} S^{de}_{RR})|^2 \} .
\end{aligned}
\label{eq:BRmue}
\end{equation}

\subsection{Rare decay $B_{s,d} \rightarrow \mu^+ \mu^-$}
\label{sec:Bmumu}

The BR of the observed decay $B_{s} \rightarrow \mu^+ \mu^-$ is consistent with the SM prediction
but due to the $\sim30\%$ uncertainty in the measurement there is still plenty
of room for NP.  For the decay $B_{d} \rightarrow \mu^+ \mu^-$ only an upper bound for its BR,
larger than the SM prediction by a factor of $\sim 7$, exists. Hence, both decays are interesting probes for NP
as the SM prediction is under fairly good control, and new, more accurate, measurements will soon be available.

In MNSUSY tree-level contributions mediated by KK-modes of neutral colorless gauge bosons
are present.  Although the photon 0-mode (\ie \ the SM photon) couples identically
to LH and RH states, this is not necessarily the case for higher KK-modes, as discussed in
Section~\ref{sec:explanationTree}, and therefore they can give contributions even though
$B_{s,d}$ are pseudoscalar mesons.
The relevant effective operators in the context of MNSUSY are
\begin{equation}
Q_{10}^{q} = (\bar b_L \gamma^\mu q_L) (\bar \mu \gamma_\mu \gamma^5 \mu)
		\qquad \tilde Q_{10}^{q} = Q_{10}^{q} (L \leftrightarrow R) \qquad q=s,d
\label{eq:effBmumu}
\end{equation}
whereas in the SM only $Q_{10}^{q}$ needs to be taken into account.
RG effects can be neglected since $Q_{10}^{q}$ and $\tilde Q_{10}^{q}$ have vanishing anomalous dimension in QCD.

The BR for the process $B_{s,d} \rightarrow \mu^+ \mu^-$, taking into account the leading SM and NP contributions,
can be written as a sum of the SM, NP and interference between the SM and NP diagrams:
$\BR_{B_q \rightarrow \mu \mu} \equiv \BR_q = \BR_q^{\rm SM} + \BR_q^{\NP} + \BR_q^{\inter}$ for $q=s,d$,
where,
\begin{equation}
	\frac{\BR_q^{\NP}}{\BR_q^{\rm SM}} =
		\frac{|C_{10}^{q \NP} - \tilde C_{10}^{q \NP}|^2}{|C_{10}^{q \SM}|^2}~~{\rm and}~~
	\frac{\BR_q^\inter}{\BR_q^\SM} =
		\frac{2 {\bf Re} \{ C_{10}^{q \SM *} (C_{10}^{q \NP} - \tilde C_{10}^{q \NP}) \} }{|C_{10}^{q \SM}|^2} .
\label{eq:BmumuNP}
\end{equation}
Here $C^q_{10}$ and $\tilde C^q_{10}$ are the coefficients of the operators in Eq.(\ref{eq:effBmumu}).
For $B_{s} \rightarrow \mu^+ \mu^-$ these are, at scale $M=\OO(1/R)$ (similarly for $B_{d}$ decay after substitutions $B^d_{X32} \rightarrow B^d_{X31}$),
\begin{equation}
\begin{aligned}
	C_{10}^{s \rm NP} (M) &= \frac{\pi^2}{3} \frac{1}{(1/R)^2} B^d_{L32} ( - B^e_{L22} S_{LL}^{de} + B^e_{R22} S_{LR}^{de} ) \\
	\tilde C_{10}^{s \rm NP} (M) &= \frac{\pi^2}{3} \frac{1}{(1/R)^2} B^d_{R32} ( - B^e_{L22} S_{RL}^{de} + B^e_{R22} S_{RR}^{de} )
\end{aligned}
\label{eq:coeffBmumu}
\end{equation}
where the coupling factors, $S_{LL}^{de}$ \etc, are as in Eq.(\ref{eq:BRmue}).

\subsection{Contributions from higher dimensional operators}
\label{sec:UVcontributions}

So far, we have computed the FCNC effects of tree-level exchange of neutral gauge boson KK-modes.
However, since MNSUSY is an effective theory valid up to a cutoff $M_5$, higher dimensional operators
arising in the UV theory with coefficients suppressed by powers of $M_5$ are in general present. 
In the spirit of MFV \cite{MFV}, we will now estimate those effects following the assumption that the only sources of
flavor violation at scale $M_5$ are the same as those in the IR, and which we treat as spurions
(transforming appropriately under the flavor symmetry).
Whereas in MFV-like theories Yukawa matrices are the sole spurions,
in MNSUSY the localization matrices $A^\Phi$ of Eq.(\ref{eq:Amatrix3rdgen}) and the matrices $\Delta^\Phi$
discussed in section~\ref{sec:statusSUSYflavor} also break flavor.
Finally, since at $M_5$ the theory becomes strongly coupled, we estimate the size of the
coefficients of the relevant flavor changing operators using NDA as
applied to theories with branes (see, \eg \ Section 3.2 of \cite{NDA}).

A particular class of processes are 4-fermion $\Delta F=1$ interactions. For example, in the case of $\mu \rightarrow \bar e e e$,
the most dangerous operators are (suppressing the $\gamma$ matrix structure)
\begin{equation}
	\lang _4 \sim \left\{
	\begin{array}{l} 
		\frac{1}{M_5^2} \frac{36 \pi^2}{(M_5 R)^2} \int d^4 \theta \ (L^\dagger S^L L)_{12} (L^\dagger L \ {\rm or} \ {\bar E}^\dagger {\bar E})_{11}\\ \\
		\frac{1}{M_5^2} \frac{36 \pi^2}{(M_5 R)^2} \int d^4 \theta \ ({\bar E}^\dagger S^{\bar E} {\bar E})_{12} (L^\dagger L \ {\rm or} \ {\bar E}^\dagger {\bar E})_{11}
	\end{array} \right.
\label{eq:NDAmu3e}
\end{equation}
where the coefficient is the NDA factor after canonically normalising the fields and $S^{L, {\bar E}}$ denote
the appropriate flavor matrices. The most dangerous spurions (in the physical basis where the lepton Yukawa matrix is diagonal) are,
respectively, $S^L = B^e_L$  and $S^{\bar E} = (B^e_R)^*$.  (Note that possible contributions from the $\Delta$ spurions are parametrically
small $\sim c^{L, {\bar E}}_{12} / (\pi N^3) \sim 10^{-4} c^{L, {\bar E}}_{12}$ compared
to that from $B^e_L$ or $(B^e_R)^*$.  In other words the contributions from off-diagonal terms
in the slepton mass-squared matrices are naturally suppressed and subdominant, unlike in most supersymmetric 
theories where they are the main source of flavor violation.) Using $M_5 = N/R$ the NDA estimate for the contribution to the BR is 
\begin{equation}
	\BR_{\mu \rightarrow \bar e e e} = \frac{\Gamma_{\mu \rightarrow \bar e e e}}{\Gamma_\mu^{\rm total}} \sim
		\frac{m_\mu^5/(64 \pi^3) |c^{\rm NDA}_{\mu \rightarrow \bar e e e}|^2}{m_\mu^5/(192 \pi^3) G_F^2} 
		 \propto \frac{v^4}{(1/R)^4} \frac{1}{N^8} |S^{L, {\bar E}}_{12}|^2 .
\label{eq:NDAmu3eBR}
\end{equation}
For $1/R=4$ TeV and $N \sim 10$, Eq.(\ref{eq:NDAmu3eBR}) results in a contribution $\BR_{\mu \rightarrow \bar e e e} \sim 10^{-7} |S^{L, {\bar E}}_{12}|^2$.
The result for the BR of the process of $\mu - e$ conversion results in a similar contribution.

When NDA-estimating the coefficient of the relevant operators for the decays $B_{s,d}\rightarrow\mu^+\mu^-$ one has to be more careful, since
now one of the fields involved is a 4D brane-localized field instead of a bulk state. In this case, the size of the coefficient is given by
\begin{equation}
	c^{\rm NDA}_{B_{s,d} \rightarrow \mu \mu} \sim \frac{1}{M_5^2} \frac{(24 \pi^2)^{3/2}}{4 \pi (M_5 R)^{3/2}} S^{Q, {\bar D}}_{32}
		= \frac{1}{(1/R)^2} \frac{(24 \pi^2)^{3/2}}{4 \pi N^{7/2}} S^{Q, {\bar D}}_{32}
\end{equation}
and for $1/R=4$ TeV and $N \sim 10$ this gives $\BR_{B_{s,d} \rightarrow \mu \mu} \sim 10^{-9} |S^{Q, {\bar D}}_{32}|^2$.
As before, we can have $S^Q = B^d_L, \Delta^Q$ and $S^{\bar D} = (B^{d}_R)^*, \Delta^{\bar D}$, with the $B$'s
giving the leading contribution.

Turning to the $\Delta F=2$ processes the effective operators would have the following structure (\eg \ for the case of kaon oscillations):
\begin{equation}
	\lang _4 \sim \left\{
	\begin{array}{l}
		\frac{M_5^4}{16 \pi^2} \int dy \ \delta(y) \int d^4 \hat \theta \ (\hat Q^\dagger S^Q \hat Q)_{21}
											(\hat Q^\dagger S^Q \hat Q \ {\rm or} \ \hat {\bar{D}}^\dagger S^{\bar D} \hat {\bar D})_{21}\\ \\
		\frac{M_5^4}{16 \pi^2} \int dy \ \delta(y) \int d^4 \hat \theta \ (\hat {\bar D}^\dagger S^{\bar D} \hat {\bar D})_{21}
											(\hat Q^\dagger S^Q \hat Q \ {\rm or} \ \hat {\bar{D}}^\dagger S^{\bar D} \hat {\bar D})_{21}
	\end{array} \right.
\label{eq:NDAKmixing}
\end{equation}
with $S^Q$ and $S^{\bar D}$ as in the previous case.
After appropriately 4D-normalizing the different fields, the size of the coefficient of these operators is
\begin{equation}
	c^{\rm NDA}_{K} \sim \frac{1}{M_5^2} \frac{36 \pi^2}{(M_5 R)^2} (S^{Q, {\bar D}}_{12})^2
			= \frac{1}{(1/R)^2} \frac{36 \pi^2}{N^4} (S^{Q, {\bar D}}_{12})^2
\label{eq:NDAKCoeff}
\end{equation} 
which, for $N\sim10$ and $1/R=4$ TeV, results in $\Delta M_K \sim 10^{-10} (S^{Q, {\bar D}}_{12})^2$ GeV and a maximum contribution
to $|\epsilon_K|$ of size $\sim 10^4 (S^{Q, {\bar D}}_{12})^2$.
The situation for $B$-meson mixing is slightly different because now two of the fields involved are brane-fields, which results in the
corresponding coefficient being
\begin{equation}
	c^{\rm NDA}_{B_{s}} \sim \frac{1}{M_5^2} \frac{24 \pi^2}{M_5 R} (S^{Q, {\bar D}}_{32})^2= \frac{1}{(1/R)^2} \frac{24 \pi^2}{N^3} (S^{Q, {\bar D}}_{32})^2
\end{equation}
and a similar coefficient $c^{\rm NDA}_{B_{d}}$ follows after replacing $S^{Q, {\bar D}}_{32}$ by $S^{Q, {\bar D}}_{31}$.
For $N\sim10$ and $1/R=4$ TeV this results in $\Delta M_{B_{s}} \sim 10^{-9} (S^{Q, {\bar D}}_{32})^2$ GeV and
$\Delta M_{B_{d}} \sim 10^{-9} (S^{Q, {\bar D}}_{31})^2$ GeV.


\section{Consequences for MNSUSY Model Building}
\label{sec:consequences}

As discussed in Sections~\ref{sec:explanationTree} and \ref{sec:limits} the size and structure of the dominant tree-level KK-mediated flavor violation 
depends upon the six matrices $B^u_{L,R}$, $B^d_{L,R}$ and $B^e_{L,R}$, which are in turn built from the five localization matrices
$A^{(Q,{\bar U},{\bar D},L,{\bar E})}$ and the six weak-to-mass eigenbasis rotation matrices $R^{u,d,e}_{L,R}$.   In the basic MNSUSY
model of Ref.\cite{MNSUSY} the full 3rd family is brane localized while the 1st and 2nd families are in the bulk, and consequently,
all five $A$ matrices have the form $A={\rm diag}(0,0,1)$.   Such a pattern of localization is not mandatory.

To understand what other patterns are possible, we recall some further aspects of MNSUSY model building \cite{MNSUSY}.  First, it is essential
for the success of EWSB with low fine-tuning that the LH and RH stop states are parametrically light compared to the gluino.  This
mandates that the 3rd-family chiral multiplets $Q_{3}$ and $\bar U_{3}$ must be brane-localized while the $SU(3)_C$ gauge multiplet must propagate in the bulk.
With this pattern of localization the gluino sees direct SSSB, acquiring a Dirac mass of size $1/(2R)$, while the stop states only see SUSY breaking
at 1-loop order.  In addition, Ref.\cite{MNSUSY} simply realised a so-called `natural' SUSY spectrum with the 1st and 2nd family sfermion states heavy by placing
these states in the bulk, so leading to them acquiring direct SSSB mass-squareds of size $1/(2R)^2$.  This feature somewhat helps fine tuning as it reduces the number of light
colored states which can be produced at the LHC, and thus allows for a reduced scale $1/R$.  In principle, however, some 3rd-family states (apart from $Q_3$ and $\bar U_{3}$) can propagate in the bulk, or, alternatively some of the 1st- and 2nd-family states can be brane-localized.  

At this point it is important to realise the second major constraint on MNSUSY model-building arising from
naturalness of EWSB: The trace of all gauged U(1) generators, in particular hypercharge, $Y$, must vanish when evaluated on the brane-localized states.  If this
is not the case then a Fayet-Illiopoulos (FI) term quadratically sensitive to the cutoff $M_5$ can arise \cite{Ghilencea:2001bw,Marti2002FI,Barbieri:2002ic}, which in
turn feeds into the Higgs soft-mass and destabilises EWSB \cite{Marti2002FI}.
Thus we are only allowed to move combinations of fields with ${\rm tr}(Y)=0$ on and
off the brane.\footnote{In the model
of Ref.\cite{MNSUSY} there was a further gauged $U(1)'$ symmetry beyond hypercharge, and tracelessness of this also imposed constraints. The presence of
$U(1)'$ was required solely to raise the Higgs mass to its observed value, but, in Ref.\cite{moreMNSUSY} it is demonstrated that
the Higgs mass is simply and elegantly achievable without the need for $U(1)'$ or other Abelian gauge symmetries.  We thus impose only the
$U(1)_Y$ FI constraints.}

This still allows a variety of structures of localization.  In addition to the basic brane-localized `3rd family' pattern already discussed there are a number of other
options with variant or enhanced flavor symmetry structure which we now enumerate:\footnote{Baroque arrangements are possible. Here we concentrate
on the simplest, most symmetrical possibilities. The localization patterns can result from an underlying bulk gauge symmetry, so explaining the (accurate)
equality of the 1st and 2nd generation bulk wavefunction profiles, see \eg \ \cite{Hebecker:2002re,Hardy:2013uxa}.}
\begin{itemize}
\item 
{\bf  `$T_3$ only':} The states ${\bar F}_3=(\bar D_{3}, L_3)$ in the 3rd-family ${\bar {\bf 5}}$ (in $SU(5)$ notation) may propagate in the bulk together with the full 1st- and 2nd-families, while only the
states $T_3=(Q_{3}, \bar U_{3}, \bar E_{3})$ making up a single ${\bf 10}$ of $SU(5)$ are localized on the brane.
In this case $A^{{\bar D},L}={\rm diag}(0,0,0)$ while
$A^{Q,{\bar U},{\bar E}} ={\rm diag}(0,0,1)$, and it is sensible to impose a $SU(3)_L \times SU(3)_{\bar D}$ flavor symmetry on all couplings which is broken (explicitly)
by the Yukawa couplings of the down-quarks and leptons, leading to a variant MFV-like scenario for these states.  This case has the feature that
the number of parametrically light sfermions is reduced compared to the basic case, so EWSB remains maximally natural at a fine-tune of only $50\%$ as in Ref.\cite{MNSUSY}.
\item
{\bf  `Ten-five split':} All three families, $T_i$, of ${\bf 10}$'s are brane localized, while all three families, ${\bar F}_i$, of ${\bar {\bf 5}}$'s propagate in the bulk.  In this case
$A^{{\bar D},L}={\rm diag}(0,0,0)$ while $A^{Q,{\bar U},{\bar E}} ={\rm diag}(1,1,1)$, and since both matrices are proportional to the unit matrix in this scenario {\it no tree-level
KK-mediated flavor violation is present}.  This exceptionally KK-flavor-violation safe and symmetrical scenario comes at a (relatively small) price of an increased
EWSB tuning $\sim 15\%$~\cite{moreMNSUSY} because of the marginally stronger LHC limits on the increased multiplicity of parametrically light squarks.   Note that it optionally allows for the imposition of a complete $U(3)^5$ flavor symmetry only broken by the Yukawa couplings, so this is a full MFV scenario.
\item
{\bf `Quark-lepton split':} All three families of quark multiplets $(Q_{i},\bar U_{i},\bar D_{i})$ are brane localized while all three families of lepton (hyper-)multiplets $(L_i, \bar E_{i})$ propagate in the bulk.
Now $A^{L, {\bar E}}={\rm diag}(0,0,0)$ while $A^{Q,{\bar U},{\bar D}} ={\rm diag}(1,1,1)$, and again no tree-level KK-mediated flavor violation is present at a price of marginally increased EW tuning, $\sim 15\%$ \cite{moreMNSUSY}.  It optionally allows for the imposition of a complete
$U(3)^5$ flavor symmetry only broken by the Yukawa couplings, so can be fully MFV.  
\end{itemize}

Having outlined the simplest MNSUSY variants we now discuss the consequences of these four arrangements for rare flavor-violating
observables.  Since, in the `ten-five split' and `quark-lepton split' patterns, KK-mediated tree-level flavor violation is
absent, we focus upon the  `3rd family' and `$T_3$ only' variants which are the most natural EWSB models of all.


\subsection{Quark sector}
\label{sec:conseqQuarks}

From the processes analysed in Section~\ref{sec:limits} we learn that in the quark sector the observables $\Delta M_K$ and $|\epsilon_K |$
bound the $\{21\}$ entries of the matrices $B^d_{L,R}$ defined in Eq.(\ref{eq:mixingMatrix}).    Generally, the limits on the entries in the
case $B^d_{L}\sim B^d_{R}$ are about 30 times stronger than if either LH or RH flavor violation can be neglected or is absent.

\begin{figure}[h]
  \begin{minipage}{0.48\linewidth}
    \centering
    \includegraphics[scale=1]{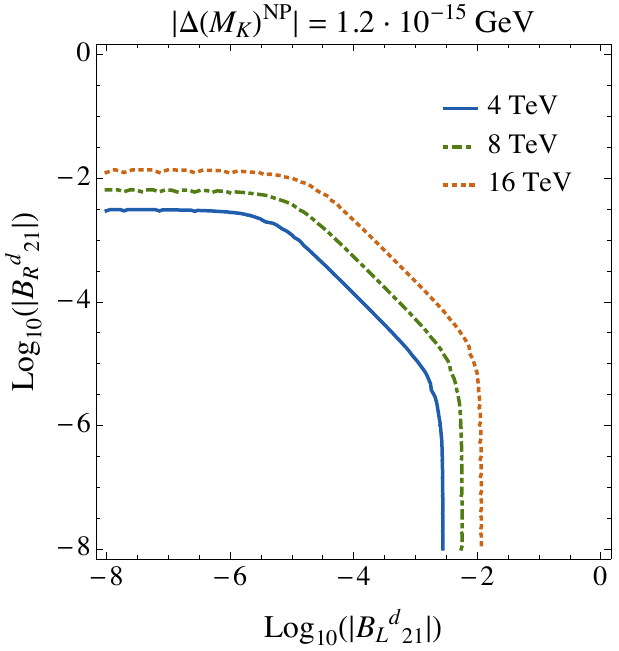}
  \end{minipage}
  \hspace{0.4 cm}
  \begin{minipage}{0.48\linewidth}
    \centering
    \includegraphics[scale=1]{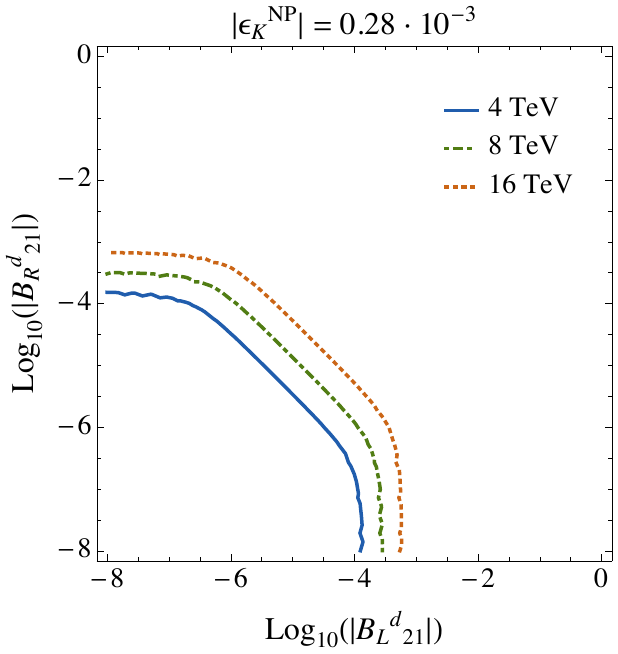}
  \end{minipage}
\caption{{\bf Left:} Limits from $|\Delta M_K |$. 
		{\bf Right:} Limits from $|\epsilon_K|$ assuming maximal phase.}
\label{fig:KMesons_Limits}
\end{figure}

\begin{figure}[h]
  \begin{minipage}{0.48\linewidth}
    \centering
    \includegraphics[scale=1]{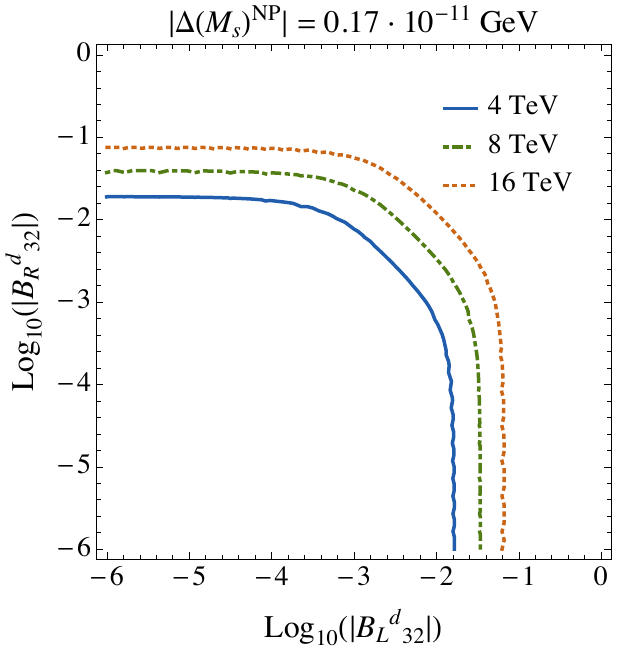}
  \end{minipage}
  \hspace{0.4 cm}
  \begin{minipage}{0.48\linewidth}
    \centering
    \includegraphics[scale=1]{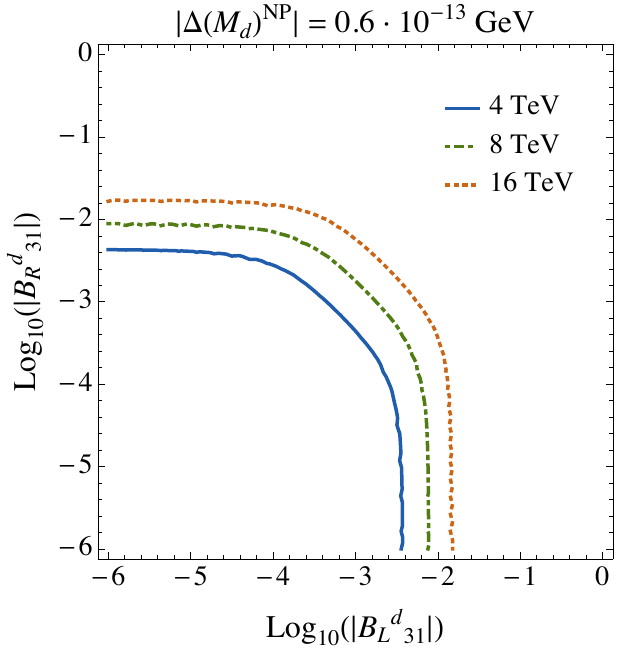}
  \end{minipage}
\caption{{\bf Left:} Limits arising from $|\Delta M_{s}|$. 
		{\bf Right:} Limits from $|\Delta M_{d}|$.}
\label{fig:BMesons_Limits}
\end{figure}

In Figure~\ref{fig:KMesons_Limits} we show the limits on $B^d_{L21}$ and $B^d_{R21}$
for NP contributions bounded by the uncertainties in the SM predictions:
$|\Delta M^{\rm NP}_K| \leq 1.2 \cdot 10^{-15} \ \GeV$ and $|\epsilon^{\rm NP}_K| \leq 0.28 \cdot 10^{-3}$.
Relative phases of terms are taken so as to give the strongest limits.
The limits from $|\epsilon_K|$ are the most stringent, as expected.
Similarly, the bounds on $\Delta M_{B_{s,d}}$ imply limits on $B^d_{X32}$ and $B^d_{X31}$. These are shown in Figure~\ref{fig:BMesons_Limits} where 
again it is clear that the case with simultaneous LH and RH flavor violation is most constrained.

We must now assess if these bounds impose severe, or relatively weak, constraints on the `3rd family' and `$T_3$ only' models. 
Recalling Eq.(\ref{eq:mixingMatrix}), the KK flavor-violation matrices are defined as $B^d_L \equiv R^{d \dagger}_L A^Q R^d_L$
and $B^d_R \equiv R^{d \dagger}_R A^{\bar D} R^d_R$ and so depend on both the basic flavor-localization matrices $A^{Q,{\bar D}}$ and
the weak-to-mass eigenbasis rotation matrices $R^d_X$.   
For both the `3rd family' and `$T_3$ only' models, $A^Q={\rm diag}(0,0,1)$, and as $R^{u \dagger}_L R^d_L = \CKM$, it is not unreasonable
to take $R^d_L$ to have a structure similar to that of $\CKM$. Specifically we assume
\begin{equation}
	(R^d_{L})_{ii} = (\CKM)_{ii} \qquad {\rm and} \qquad (R^d_{L})_{ij} = y (\CKM)_{ij} \ {\rm for} \ i \neq j 
\label{eq:RdLassumptions}
\end{equation}
where $y$ is an, a priori $\OO(1)$, free parameter to be constrained. This implies that for both variants the entries of the mixing matrix
$B^d_L$ take the form
\begin{equation}
	B^d_{L21} = |y|^2 V_{ts}^* V_{td} \sim 3\cdot 10^{-4},~~
	B^d_{L32} = y V_{tb}^* V_{ts} \sim 3\cdot 10^{-2},~~
	B^d_{L31} = y V_{tb}^* V_{td} \sim 8\cdot 10^{-3}.
\label{eq:BdLstructure}
\end{equation} 
Note that Eq.(\ref{eq:RdLassumptions}) implies that the off-diagonal entries of $B^d_L$ are {\it naturally small},
so that LH KK-mediated flavor violation is suppressed even in the basic `3rd family' model.

Turning to $B^d_R$, the MNSUSY variants now differ: In the `3rd family' case $A^{\bar D}={\rm diag}(0,0,1)$, while in the `$T_3$ only' case
$A^{\bar D}={\rm diag}(0,0,0)$.  This last fact then implies $B^d_R=0$ for the `$T_3$ only' variant irrespective of the rotation matrices.  
For the `3rd family' case $B^d_R=R^{d \dagger}_R A^{\bar D} R^d_R \neq 0$ and the simplest assumption to make regarding the RH rotation matrices
is that they have the same hierarchical structure as the LH matrices, giving $B^d_R$ approximately equal to $B^d_L$ of Eq.(\ref{eq:BdLstructure}). 
Thus, with not unreasonable assumptions, the two variants, `3rd family' and `$T_3$ only', respectively realise
the $B^d_L \sim B^d_R$ and $B^d_R=0$ situations.

\begin{figure}[h]
  \begin{minipage}{0.48\linewidth}
    \centering
    \includegraphics[scale=1]{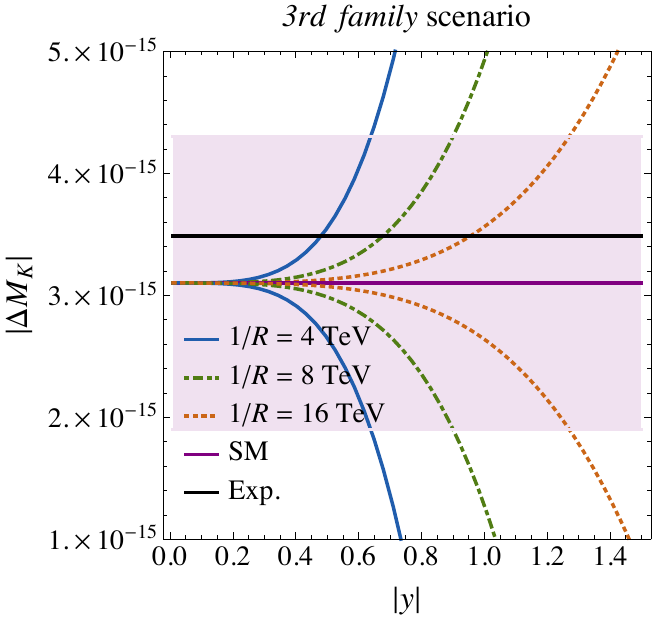}
  \end{minipage}
  \hspace{0.4 cm}
  \begin{minipage}{0.48\linewidth}
    \centering
    \includegraphics[scale=1]{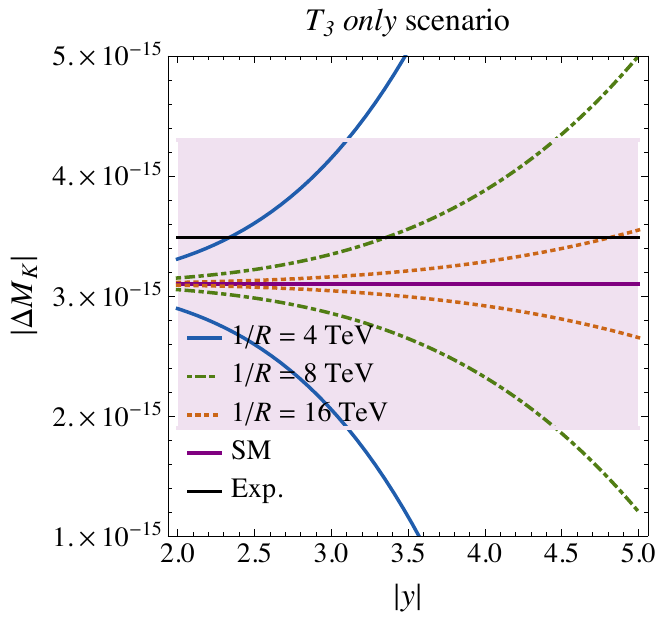}
  \end{minipage}
\caption{Predictions for $|\Delta M_K^{\rm SM}| \pm |\Delta M_K^{\rm NP}|$
	in the `3rd family' (left) and `$T_3$ only' (right) cases as a function of $|y|$
	assuming Eq.(\ref{eq:RdLassumptions}).
	Purple line is the SM prediction and its uncertainty (shaded area).
	Black line is the measurement with experimental error (dashed).}
\label{fig:DeltaMKconsequences}
\end{figure}

\begin{figure}[h]
  \begin{minipage}{0.48\linewidth}
    \centering
    \includegraphics[scale=1]{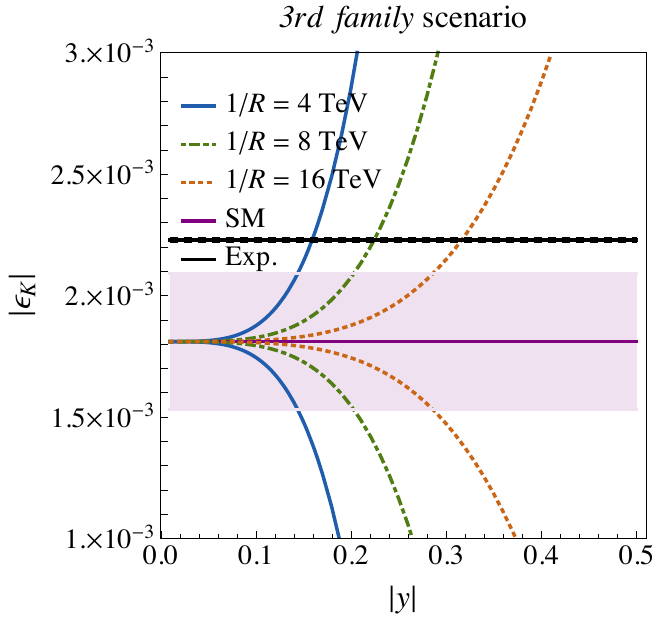}
    \label{fig:epsilonK_vs_BdL21}
  \end{minipage}
  \hspace{0.4 cm}
  \begin{minipage}{0.48\linewidth}
    \centering
    \includegraphics[scale=1]{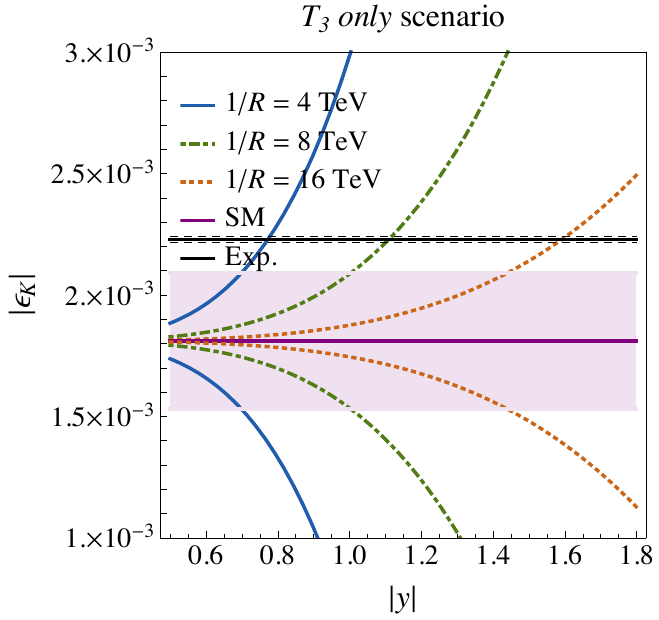}
    \label{fig:epsilonK_vs_y}
  \end{minipage}
\caption{Predictions for $|\epsilon_K^{\rm SM}| \pm |\epsilon_K^{\rm NP}|$
	in the `3rd family' (left) and `$T_3$ only' (right) cases as a function $|y|$
	assuming Eq.(\ref{eq:RdLassumptions}).
	Legends as in Figure~\ref{fig:DeltaMKconsequences}.}
\label{fig:epsilonKconsequences}
\end{figure}

\begin{figure}[h]
  \begin{minipage}{0.48\linewidth}
    \centering
    \includegraphics[scale=1]{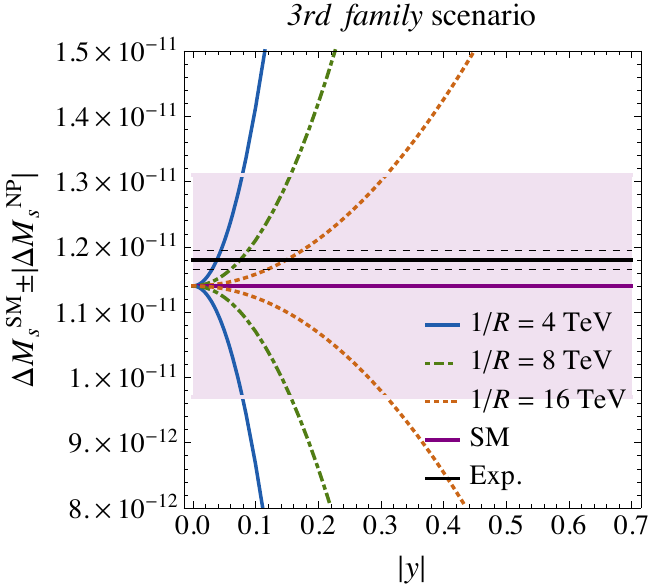}
  \end{minipage}
  \hspace{0.4 cm}
  \begin{minipage}{0.48\linewidth}
    \centering
    \includegraphics[scale=1]{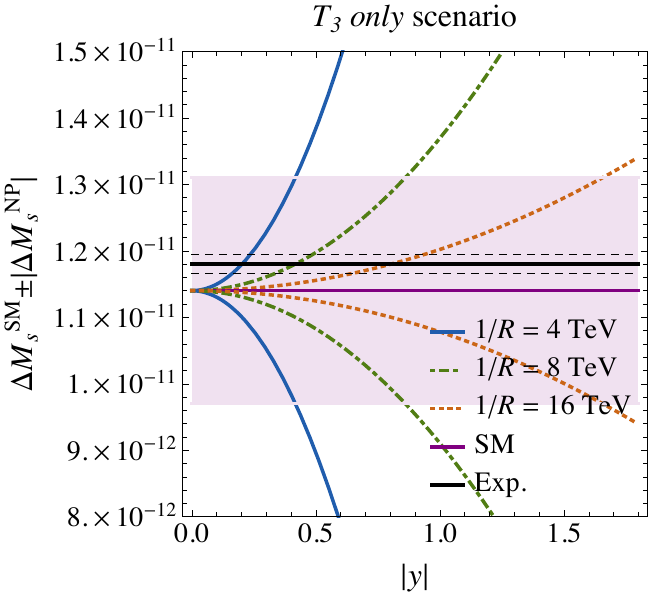}
  \end{minipage}
	\caption{Predictions for $|\Delta M_{s}^{\rm SM}| \pm |\Delta M_{s}^{\rm NP}|$
	in the `3rd family' (left) and `$T_3$ only' (right) cases
	as a function of $|y|$ assuming Eq.(\ref{eq:RdLassumptions}).
	Legends as in Figure~\ref{fig:DeltaMKconsequences}.}
\label{fig:DeltaMsconsequences}
\end{figure}

\begin{figure}[h]
  \begin{minipage}{0.48\linewidth}
    \centering
    \includegraphics[scale=1]{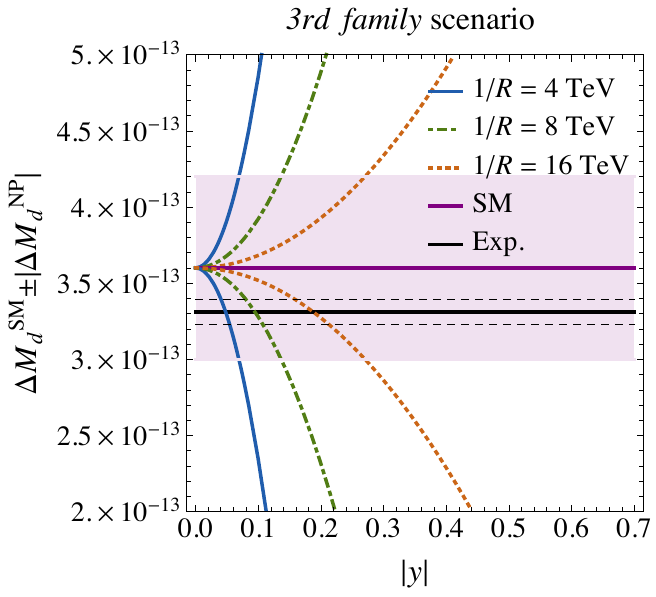}
  \end{minipage}
  \hspace{0.4 cm}
  \begin{minipage}{0.48\linewidth}
    \centering
    \includegraphics[scale=1]{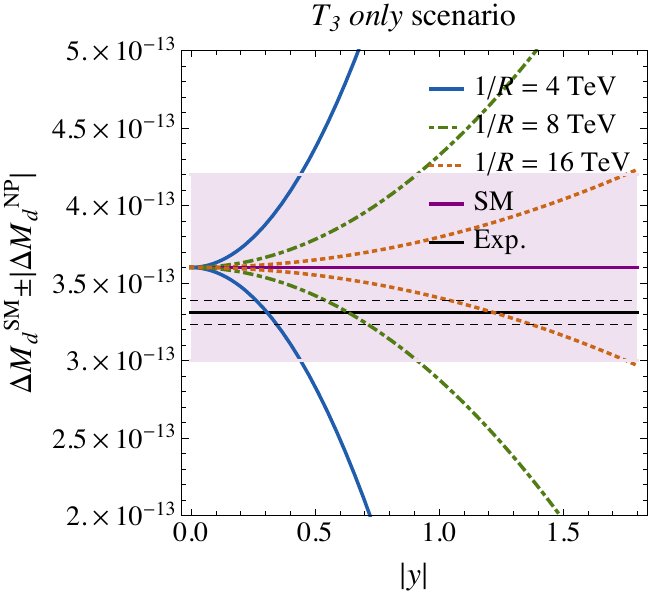}
  \end{minipage}
	\caption{Predictions for $|\Delta M_{d}^{\rm SM}| \pm |\Delta M_{d}^{\rm NP}|$
	in the `3rd family' (left) and `$T_3$ only' (right) cases
	as a function of $|y|$ assuming Eq.(\ref{eq:RdLassumptions}).
	Legends as in Figure~\ref{fig:DeltaMKconsequences}.}
\label{fig:DeltaMdconsequences}
\end{figure}

Making the assumption in Eq.(\ref{eq:RdLassumptions}),
Figures~\ref{fig:DeltaMKconsequences} and \ref{fig:epsilonKconsequences} show the theoretical prediction for
$\Delta M_K$ and $|\epsilon_K|$ for several values of $1/R$ as a function of $|y|$ in the `3rd family' (left) and `$T_3$ only' (right) scenarios.
Similarly, Figures~\ref{fig:DeltaMsconsequences} and \ref{fig:DeltaMdconsequences} show the theoretical prediction for $\Delta M_{s,d}$
for several values of $1/R$ as a function of $|y|$.



From the left panels of Figures~\ref{fig:DeltaMKconsequences}-\ref{fig:DeltaMdconsequences}, we see that the basic `3rd family'
scenario requires a reduction of the off-diagonal elements of
the eigenbasis rotation matrices beyond that naturally provided by Eq.(\ref{eq:RdLassumptions}).
The strongest constraints arise from $|\Delta M_{s,d}|$ in the $B$-system, with $|y|\lsim 0.08$ at $1/R\sim 4$ TeV,
relaxing to $|y|\lsim 0.3$ at $1/R\sim 16$ TeV, with $|\epsilon_K|$ providing the next strongest constraint. 
Although not as severe as might have naively been feared, this constraint motivates turning to the `$T_3$ only' scenario.  

As can be seen from the right panels of Figures~\ref{fig:DeltaMKconsequences}-\ref{fig:DeltaMdconsequences}, in the `$T_3$ only' scenario, the constraint
is now at worst $|y|\lsim 0.5$ at $1/R\sim 4$ TeV, so essentially
no tuning at all given that $R^{u \dagger}_L R^d_L = \CKM$ easily allows $y\sim 0.5$ in Eq.(\ref{eq:RdLassumptions}) if LH rotations
supply a comparable contribution as the RH rotations to $\CKM$.   We also learn that for the `$T_3$ only' case, unless the compactification
scale $1/R$ is significantly larger than 4 TeV (thus worsening EWSB tuning to $\lsim 10\%$), {\it if
the accuracy of the theoretical SM predictions are improved, deviations in rare flavor observables
should be seen, especially in the $B$-meson system}.\footnote{If nature chose  $R^d_L={\mathbb{I}}_3$ and $R^u_L = \CKM$, then
there would be
no effect in the down-type quark sector.}

\subsection{Lepton sector}
\label{sec:conseqLeptons}

\begin{figure}[h]
  \begin{minipage}{0.48\linewidth}
    \centering
    \includegraphics[scale=1]{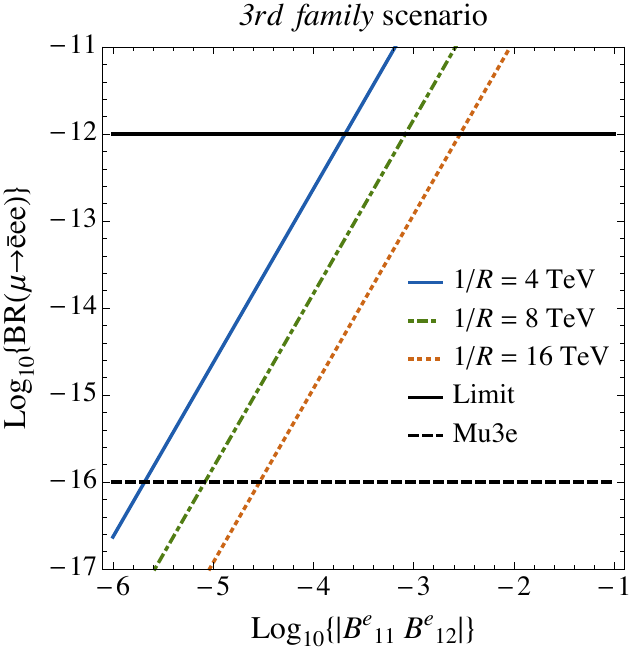}
  \end{minipage}
  \hspace{0.4 cm}
  \begin{minipage}{0.48\linewidth}
    \centering
    \includegraphics[scale=1]{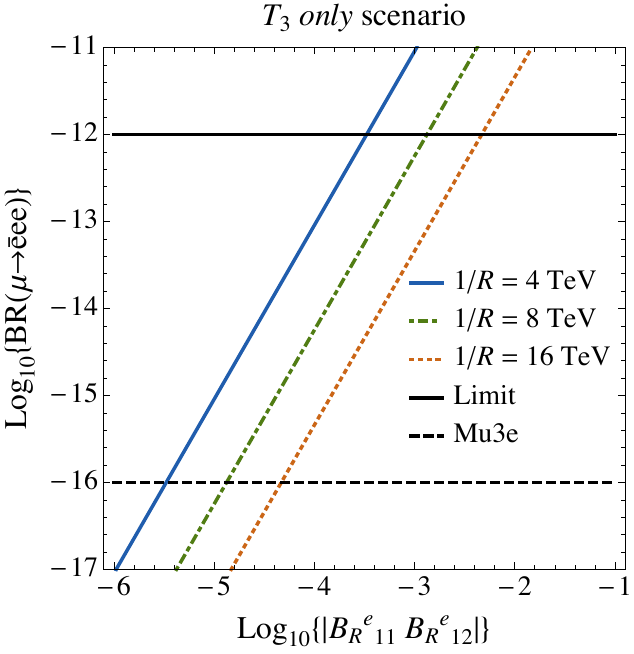}
  \end{minipage}
\caption{Predictions for $\BR(\mu \rightarrow {\bar e}ee)$ in `3rd family' case taking $B^e_R=B^e_L=B^e$ (left)
	and in the `$T_3$ only' case (right).
	Black solid line is experimental bound, dashed line is projected Mu3e sensitivity.}
    \label{fig:mu3e_limits}
\end{figure}

In the lepton sector the structures of the weak-to-mass eigenbasis rotation matrices are even less constrained by experiment than in the quark case.
As a working assumption in the `3rd family' case we take $R^e_L \sim R^e_R \equiv R^e$ and therefore $B^e_L \sim B^e_R \equiv B^e$.
Then $\BR(\mu \rightarrow {\bar e}ee)$ only depends on the product $|B^e_{11} B^e_{12}|$ and Figure~\ref{fig:mu3e_limits} (left) shows the resulting limits.
For $1/R \sim 4 \ \TeV$ the constraint is $|B^e_{11} B^e_{12}|\lsim 2\cdot10^{-4}$.   If the charged leptons were to have a hierarchical structure of
weak-to-mass eigenbasis mixings similar to that of the quarks, then this bound is not too severe as it is satisfied for e.g. off-diagonal matrix
elements $|R^e_{31}|, |R^e_{32}| = \OO(10^{-1})$.  Of course the large observed mixings in the neutrino sector must then arise from large mixings in the
singlet RH neutrino Majorana mass matrix.  Whether this is a reasonable possibility or not depends on the details of neutrino model building, a subject which is
beyond the scope of the present work.

Turning to the `$T_3$ only' variant, it leads to $B^e_L \equiv 0$ as now $L_{3}$ is in the bulk.  
The BR in this case thus only depends on $|B^e_{R11} B^e_{R12}|$, but the constraints are only marginally
weaker than in the `3rd family' scenario, as can be seen in Figure~\ref{fig:mu3e_limits} (right), so the phenomenology is very similar.

Turning now to $\mu N \to e N$ conversion it is important to realise that the master expression Eq.(\ref{eq:BRmue})
not only depends on the $\{12\}$ matrix elements of $B^e_{L,R}$ in the lepton sector, but also on the $\{11\}$ matrix
elements of $B^{u,d}_{L,R}$ in the quark sector, and these elements are {\it small} unless (some of) the 1st-family quarks are
localized on the brane as in the `Ten-five split' and `Quark-leptons split' cases.  The reason for this is simply
that before weak-to-mass eigenbasis rotations are taken into account the KK-gauge bosons that mediate flavor violation
have {\it zero} diagonal couplings to bulk fermion zero-mode states.  Explicitly one finds in the `3rd family' case
$B^{u,d}_{X11} = |R^{u,d}_{X31}|^2$ for $X=L,R$.  

\begin{figure}[h]
  \begin{minipage}{0.48\linewidth}
    \centering
    \includegraphics[scale=1]{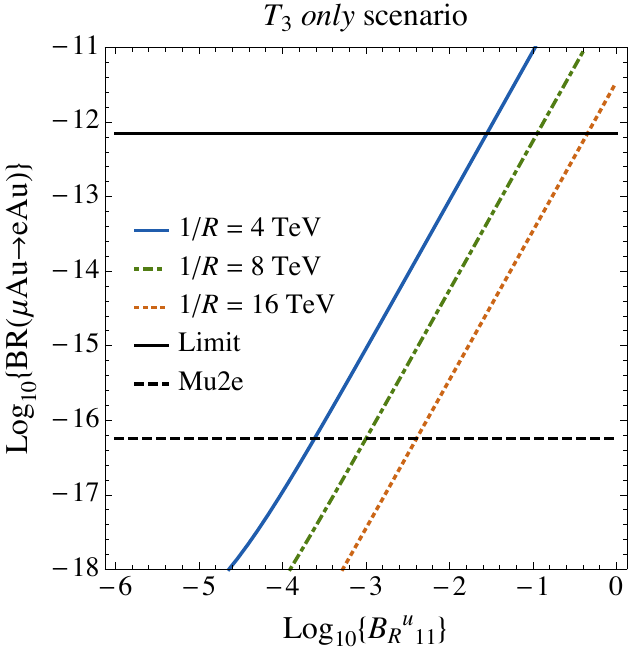}
  \end{minipage}
  \hspace{0.4 cm}
  \begin{minipage}{0.48\linewidth}
    \centering
    \includegraphics[scale=1]{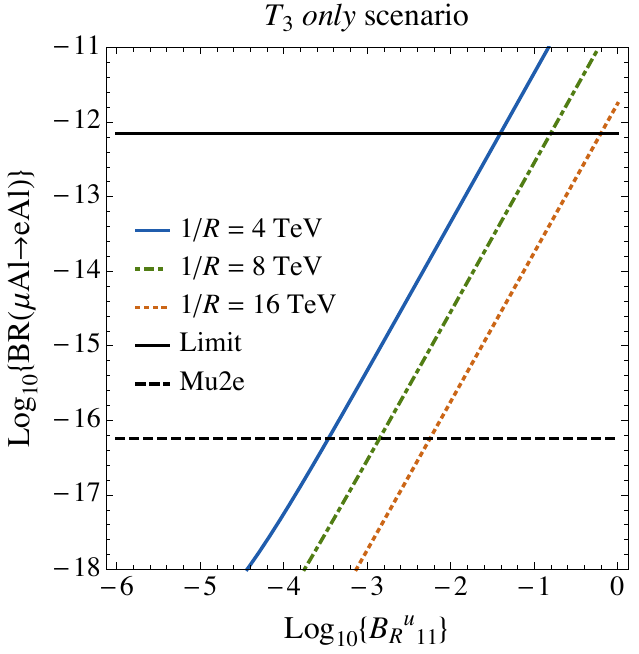}
  \end{minipage}
\caption{Value of $\BR(\mu \Au \rightarrow e \Au)$ (left) and $\BR(\mu \Al \rightarrow e \Al)$ (right)
	in the `$T_3$ only' case with assumptions as given in text.
	Black solid line is the experimental bound (Au targets),
	dashed line is the Mu3e experiment sensitivity (Al targets).}
    \label{fig:mue}
\end{figure}

Similarly, in the interesting `$T_3$ only' case $B^{u}_{X11} = |R^{u}_{X31}|^2$ for $X=L,R$, and
$B^{d}_{L11} = |R^{d}_{L31}|^2$, but both $B^d_R \equiv 0$ and $B^e_L \equiv 0$.
Given the structure of rotation matrices\footnote{Regarding $B^u_{L11}$, 
we have $B^u_L = \CKM B^d_L \CKMd$.} assumed in Eq.(\ref{eq:RdLassumptions}) we then have
$B^d_{L11} = |y|^2 |V_{td}|^2$ and $B^u_{L11} \approx B^d_{L11}$.   
Given the results discussed in Section~\ref{sec:conseqQuarks} and for $\BR(\mu \rightarrow {\bar e}ee)$,
we may assume $y\lsim 0.4$, a bound which we choose, for illustrative purposes, to saturate.  Similarly
we take $B^e_{R12} \approx 10^{-2}$ consistent with the sizes of the elements
of $B^e_{R}$ necessary to satisfy the $\BR(\mu \rightarrow {\bar e}ee)$ constraints. 
Figure~\ref{fig:mue} then shows the resulting predictions for the BR's of the processes
$\mu \Au \rightarrow e \Au$ (left) and $\mu \Al \rightarrow e \Al$ (right) as a function of $B^u_{R11}$.
For $1/R=4 \ \TeV$, $B^u_{R11}$ is constrained $\lsim 10^{-2}$ implying $|R^u_{R31}| \lsim 10^{-1}$.

Finally, we turn to the process $B_{s,d} \to \mu^+\mu^-$. In the `$T_3$ only' case, for $B_{s} \rightarrow \mu^+ \mu^-$
the Wilson coefficients are those in Eq.(\ref{eq:coeffBmumu}) after setting $B^d_R, B^e_L \equiv 0$,
in which case $\tilde C_{10}^{s \rm NP} \equiv 0$ and $C_{10}^{s \rm NP}$ only depends on the product $|B^d_{L32}| |B^e_{R22}|$.
(Similarly $B_{d}$ decays can be obtained after the substitutions $B^d_{X32} \rightarrow B^d_{X31}$.)
Taking into account the just analysed constraints from the purely hadronic or purely leptonic process
we may assume Eq.(\ref{eq:RdLassumptions}) to hold with $|y| \lsim 0.4$
and $B^e_{R22} = \frac{|B^e_{R12}|^3}{|B^e_{R11} B^e_{R12}|} \lsim 10^{-2}$.
We then find that MNSUSY contributions to $\BR(B_{s,d} \rightarrow \mu^+ \mu^-)$ are much smaller than the SM predictions.
Given the current experimental errors and theoretical uncertainties this NP contribution is therefore completely irrelevant.

\subsection{Dipole processes}
\label{sec:dipole}

Although our focus in this work has been the new tree-level FCNC processes that can occur in MNSUSY theories, here 
we wish to make a few brief comments concerning the $\mu \rightarrow e \gamma$ and $\bar B \rightarrow X_s \gamma$
processes that arise via flavor-violating dipole-like operators. We hope to return to these processes in a subsequent publication.

Of particular interest is the decay $\mu \rightarrow e \gamma$, with a current experimental upper bound
$\BR(\mu \rightarrow e \gamma)<5.7\cdot 10^{-13}$, which is far above the
SM prediction and therefore represents a clear signal of NP if observed. From 2016 an upgraded version of the MEG experiment will
start taking data with projected sensitivity of $6 \cdot 10^{-14}$ \cite{MEGupgrade2013}.

Unlike the previous processes we have considered, the leading contributions appear at one loop and are calculable.  This
is because they depend on SUSY being broken, since it is well known that dipole-like interactions vanish in the SUSY limit \cite{FerraraDipole}.
Since SUSY breaking in MNSUSY turns on at scale $1/R$, at energy scales much larger than these, like $M_5$, SUSY appears to be preserved and therefore no
UV contribution to these processes is possible.

\begin{figure}[h]
    \centering
    \includegraphics[scale=0.6]{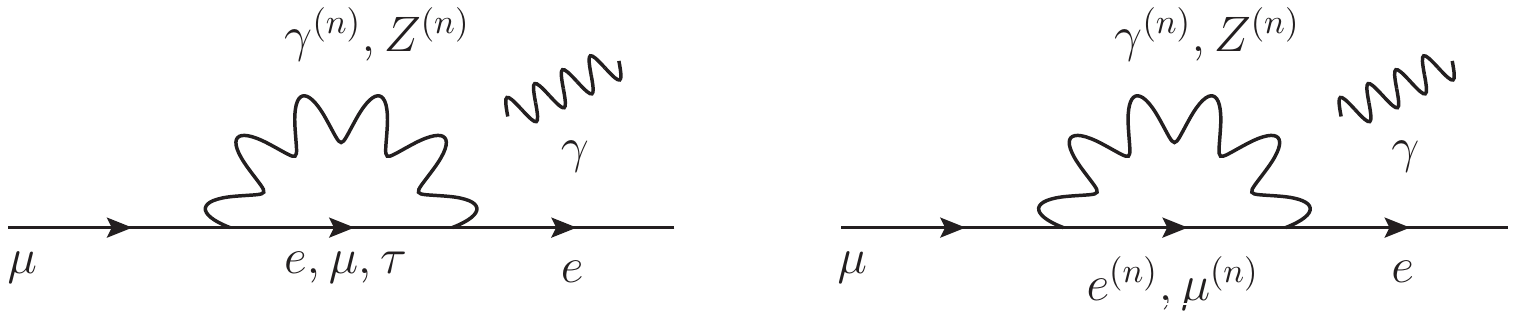}
    \caption{Diagrams contributing to the process $\mu \rightarrow e \gamma$,
	with neutral gauge boson KK-modes and charged leptons (0-modes (left) and KK-modes (right)) in the loop.
	Similar diagrams with $W$ boson KK-modes and neutrinos (0-modes and KK-modes) are also present.
	Analogous diagrams with the corresponding sparticles running in the loop also exist.}
  \label{fig:muegammanonSUSY}
\end{figure}

\begin{figure}[h]
    \centering
    \includegraphics[scale=0.6]{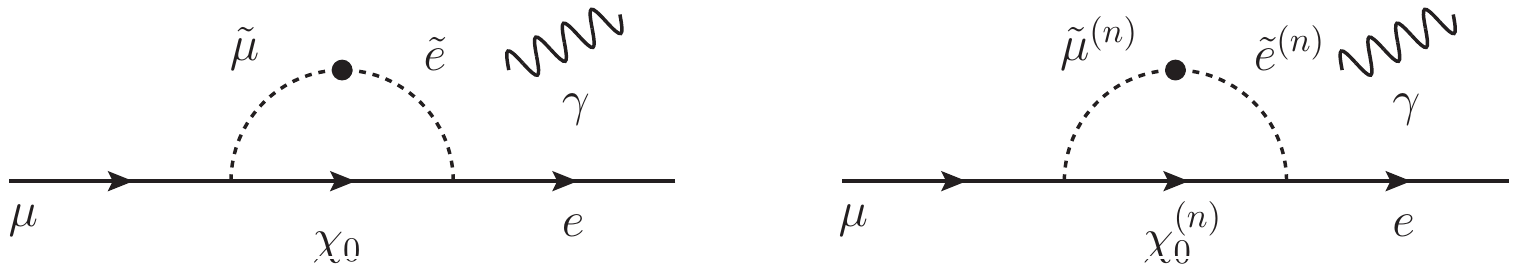}
    \caption{Diagrams contributing to the process $\mu \rightarrow e \gamma$,
	with a neutralino and charged sleptons in the loop (and also their KK-modes).
	The dot represents a mass insertion and corresponds to some off-diagonal entry of the matrices $\Delta^{L,{\bar E}}$.
	Diagrams with charginos and sneutrinos and their KK-towers are also present.}
  \label{fig:muegammaSUSY}
\end{figure}

Turning to an estimate of the IR contribution, Figures~\ref{fig:muegammanonSUSY} and \ref{fig:muegammaSUSY} show some of the one-loop diagrams contributing.
For example, those in Figure~\ref{fig:muegammanonSUSY} arise because of the 3rd family being localized on the brane and therefore would involve some entries of the
matrices $B^e_{L,R}$. On the other hand, those in Figure~\ref{fig:muegammaSUSY} involve only off-diagonal entries on the slepton mass-squared matrix
and therefore would be proportional $\Delta^{L, {\bar E}}_{12}$. It is obvious that this latter kind of diagram would vanish if SUSY was unbroken,
as in that case $\Delta^{L, {\bar E}}=0$. The diagrams in Figures~\ref{fig:muegammanonSUSY}, though, would be cancelled by the analogous
diagrams where sparticles run in the loop.

Now, we consider the operators of lowest dimension
with a dipole-like structure. Since SUSY breaking must be involved, these are of the form:
\begin{equation}
	\lang_4 \sim \left\{
	\begin{array}{l}
		\frac{\pi R}{16 \pi^2} \frac{e g^2}{(1/R)^5} \int dy \ \delta(y) \int d^4 \theta \
			H_u^{5D \dagger} ({\bar E}^{5D} \tilde S D^\alpha L^{5D})_{12} W^{5D}_\alpha X^\dagger + {\rm h.c}\\ \\
		\frac{\pi R}{16 \pi^2} \frac{e g^2}{(1/R)^5} \int dy \ \delta(y) \int d^4 \theta \
			H_u^{5D} (L^{5D \dagger} \tilde S^\prime D^\alpha {\bar E}^{5D \dagger})_{12} W^{5D}_\alpha X + {\rm h.c}
	\end{array} \right.
\end{equation}
where $\tilde S, \tilde S^\prime$ are flavor spurions and the coefficient has been estimated
taking into account that the leading contribution is a loop process and that one KK-summation is involved.\footnote{When summing over a whole
KK-tower the degree of convergence of a loop integral can be decreased (see \eg \ \cite{Csaki2010}). In our case, the diagrams have
a degree of convergence reduced by 1 compared to the standard 4D calculation and an additional numerator factor of $\OO(\pi R)$ appears in the
estimates. Roughly speaking this can be understood from the behaviour of the scalar field Euclidian propagator after summing
over a whole KK tower: $\sum_{n=0}^\infty (p_E^2 + (n/R)^2)^{-1} \sim \pi R/( 2 p_E)$ for large $p_E$.
Also note that in our case convergence of the loop integral was not an issue: we know it must be convergent since the UV limit of the
theory is supersymmetric.}

After appropriately 4D-normalizing the different fields and taking into account the relevant non-zero vev's
($\langle H_u^{(0)} \rangle = v/\sqrt{2}$, $\langle F_X \rangle \sim 1/R^2$), we get
\begin{equation}
	\lang_4 \sim \left\{
	\begin{array}{l}
		\frac{e g^2}{16 \pi} \frac{v/\sqrt{2}}{\pi^2 (1/R)^2} \tilde S_{12} (\bar e_R \sigma^{\mu \nu} \mu_L) F_{\mu \nu}  + {\rm h.c}\\ \\
		\frac{e g^2}{16 \pi} \frac{v/\sqrt{2}}{\pi^2 (1/R)^2} \tilde S^\prime_{12} (\bar e_L \sigma^{\mu \nu} \mu_R) F_{\mu \nu}  + {\rm h.c}
	\end{array} \right.
\label{eq:muegammaOperators}
\end{equation}
Since the $\tilde S$ spurions must transform in the same way as the lepton Yukawa matrix $\lambda_e$, the possible structures are
\begin{equation}
	\tilde S = \left\{
	\begin{array}{l}
		\lambda_e B^e_L, \ \lambda_e \Delta^L\\
		B^e_R \lambda_e , \ \Delta^{{\bar E} *} \lambda_e\\
		B^e_R \lambda_e B^e_L , \ \Delta^{{\bar E} *} \lambda_e \Delta^L, \ B^e_R \lambda_e \Delta_L, \ \Delta^{{\bar E} *} \lambda_e B^e_L
	\end{array} \right.
\label{eq:muegamma_spurions}
\end{equation}
and similar arrangements for $\tilde S^\prime$. Importantly one power of $\lambda_e$ must be involved.
Moreover, matrices $\tilde S$ and $\tilde S^\prime$ involving one $B$ or one $\Delta$ matrix result in a \{12\} term that can only be proportional
to $y_e$ or $y_\mu$, whereas once two of them are involved the leading contribution is proportional to $y_\tau$.
This matches our intuition if for example we consider $B$ matrices only: if both matrices $B^e_{L,R}$ are present, a chirality flip may occur in the
internal fermion line in the diagrams on the left of Figure~\ref{fig:muegammanonSUSY}, picking up a factor of $y_\tau$,
whereas if only $B^e_{R}$ exists (for example in the `$T_3$ only' scenario) then this is no longer possible and the leading contribution must be proportional to $y_\mu$.
Note also that, unlike in the SM, both operators $(\bar e_R \sigma^{\mu \nu} \mu_L) F_{\mu \nu}$ and $(\bar e_L \sigma^{\mu \nu} \mu_R) F_{\mu \nu}$
may be equally important.
A contribution proportional to $y_\tau$ may still be possible if only $B^e_R$ is present by using one of the $\Delta$ matrices, although in this case
we pay an extra suppression from the off-diagonal terms of these matrices. 

As an illustrative example, consider the cases $\tilde S = B^e_R \lambda_e$ and $\tilde S = B^e_R \lambda_e B^e_L$,
which result in $\tilde S_{12} = y_\mu B^e_{R12}$ and $\tilde S_{12} = y_\tau B^e_{L32} B^e_{R12}$ respectively.
In these cases, the contributions to the BR of this process are of order $\BR_{\mu \rightarrow e \gamma} \sim 10^{-10} |B^e_{R12}|^2$,
and $10^{-8} |B^e_{L32} B^e_{R12}|^2$ respectively. On the other hand, if $\tilde S = \Delta^{{\bar E}*} \lambda_e$ and
$\tilde S = \Delta^{{\bar E}*} \lambda_e \Delta^L$ we find $\BR_{\mu \rightarrow e \gamma} \sim 10^{-17} |c^{\bar E}_{12}|^2$, and 
$10^{-20} |c^{L}_{32} c^{\bar E}_{12}|^2$.
We can see how thanks to the natural suppression of the off-diagonal elements of the $\Delta$ matrices, these operators
are subdominant.\footnote{The anomaly mediated contribution to $A$-terms and Majorana gaugino masses gives a contribution to the BR
that is also subdominant. This can be seen by taking into account that an insertion from an $A$-term needs to appear together with a Majorana
gaugino mass factor. Defining $\delta^A \equiv \frac{A (v/\sqrt{2})}{1/(2R)^2}$
with the form of $A$ given in Eq.(\ref{eq:AMSUSYbreaking}), the contribution to the BR is like that from taking
$\tilde S = \Delta^{{\bar E} *} \lambda_e$ but with $\delta^A m_\lambda / m_\mu$ instead of $\Delta^{{\bar E} *}_{12}$.
For $F_\phi \sim 1/R =4$ TeV the contribution to the BR is $\sim 10^{-18}$.}

Note that for variant localization patterns, e.g. the `$T_3$ only'  scenario where $B^e_L \equiv 0$, the main contribution to the BR of this radiative
decay gives $\BR_{\mu \rightarrow e \gamma} \sim 10^{-10} |B^e_{R12}|^2$, which is consistent with the current upper bound but in
reach of future experiments for $|B^e_{R12}| \sim 10^{-2}$.

Finally, completely analogous considerations apply for the radiative decay $\bar B \rightarrow X_s \gamma$.
The leading contribution involving
$B$ matrices results in $\BR \sim 10^{-7} |B^d_{R23}|^2$, whereas the main contribution involving
one $\Delta$ matrix leads to $\BR \sim 10^{-13} |c^{Q, {\bar D}}_{23}|^2$.  These contributions are both
negligible compared to the present and expected future experimental precision.

\section{Conclusions}
\label{sec:conclusions}

We have found in this work that rare CP-conserving flavor violating processes (and $\epsilon_K$) provide an important window
onto the structure of the recently proposed theory of Maximally Natural Supersymmetry.  Quite generally in MNSUSY we find that
flavor violation from the off-diagonal elements in the mass-squared sfermion masses account for a sub-leading, and harmless,
contribution to rare processes, contrary to what happens in most SUSY theories and therefore solving the so-called `SUSY flavor problem'.

We find, however, that in the original MNSUSY model \cite{MNSUSY} a more important source of non-MFV flavor violation arises from the fact that
the full 3rd family of matter is localized on one of the branes, whereas the 1st and 2nd family propagate in the 5D bulk.
In this original `3rd family' scenario, an extra suppression of flavor violation is needed compared to that automatically present
in order to satisfy current constraints.

This motivates consideration of a number of alternative realizations of MNSUSY also with minimal or low fine-tuning of EWSB
(varying from $\sim 50\%$ to $\sim 15\%$), but with enhanced flavor symmetry structure
for the matter localization. For example, in the `$T_3$ only' scenario, where $Q_3$, $\bar U_3$ and $\bar E_3$ are brane localized but
the rest of the 3rd family is allowed to propagate in the 5D bulk, current constraints are naturally satisfied but,
intriguingly, flavor violation is within reach of future experiments if further suppressions are not present.

Generally speaking we find that experiments looking for flavor violation in the lepton sector (via the processes $\mu \rightarrow {\bar e}ee$,
$\mu - e$ nuclear conversion, and $\mu \rightarrow e \gamma$) should typically see signals in future upgrades.
Deviations in some observables in the quark sector are also possible if
the accuracy of the theoretical SM predictions are improved.

Overall we find that MNSUSY theories give a rich and well-motivated class of models where, depending on the exact localization
pattern, new contributions to rare flavor-violating processes can arise at an experimentally interesting level, either with a MFV structure
or with simple forms of non-MFV flavor violation.


\section*{Acknowledgements}

We thank M. Baryakhtar,  G. D'Ambrosio, E. Hardy and, especially, S. Dimopoulos, U. Haisch and K. Howe for discussions.
This work was partially supported by ERC grant BSMOXFORD no. 228169. IGG and JMR thank
the Stanford Institute for Theoretical Physics and the CERN Theory Group for hospitality during portions of this work.
IGG is financially supported by the EPSRC/STFC and a Scatcherd European Scholarship from the University of Oxford.


\bibliography{flavourReferences}

\end{document}